# Tunable optical topological transitions of plasmon polaritons in WTe$_2$ van der Waals films

Yuangang Xie[1], Chong Wang[2,3]*, Fucong Fei[4,5]*, Yuqi Li[2,3], Qiaoxia Xing[1], Shenyang Huang[1], Yuchen Lei[1], Jiasheng Zhang[1], Lei Mu[1], Yaomin Dai[6], Fengqi Song[4,5], Hugen Yan[1]*

1, State Key Laboratory of Surface Physics, Key Laboratory of Micro and Nano-Photonic Structures (Ministry of Education), and Department of Physics, Fudan University, Shanghai 200433, China.

2, Centre for Quantum Physics, Key Laboratory of Advanced Optoelectronic Quantum Architecture and Measurement (MOE), School of Physics, Beijing Institute of Technology, Beijing, 100081, China.

3, Beijing Key Lab of Nanophotonics & Ultrafine Optoelectronic Systems, School of Physics, Beijing Institute of Technology, Beijing, 100081, China.

4, National Laboratory of Solid State Microstructures, Collaborative Innovation Center of Advanced Microstructures, and College of Physics, Nanjing University, Nanjing 210093, China.

5, Atom Manufacturing Institute (AMI), Nanjing 211805, China.

6, Center for Superconducting Physics and Materials, National Laboratory of Solid State Microstructures and Department of Physics, Nanjing University, Nanjing 211805, China.

*Corresponding author. Email: chongwang@bit.edu.cn (C.W.), feifucong@nju.edu.cn (F.F.), and hgyan@fudan.edu.cn (H.Y.).

Naturally existing in-plane hyperbolic polaritons and the associated optical topological transitions, which avoid the nano-structuring to achieve hyperbolicity, can outperform their counterparts in artificial metasurfaces. Such plasmon polaritons are rare, but experimentally revealed recently in WTe$_2$ van der Waals thin films. Different from phonon polaritons, hyperbolic plasmon polaritons originate from the interplay of free carrier Drude response and interband transitions, which promise good intrinsic tunability. However, tunable in-plane hyperbolic plasmon polariton and its optical topological transition of the isofrequency contours to the elliptic topology in a natural material have not been realized. Here we demonstrate the tuning of the optical topological transition through Mo-doping and temperature. The optical topological transition energy is tuned over a wide range, with frequencies ranging from 429 cm$^{-1}$ (23.3 microns) for pure WTe$_2$ to 270 cm$^{-1}$ (37.0 microns) at the 50% Mo-doping level at 10 K. Moreover, the temperature-induced blueshift of the optical topological transition energy is also revealed, enabling active and reversible tuning. Surprisingly, the localized surface plasmon resonance in skew ribbons shows unusual polarization dependence, accurately manifesting its topology, which renders a reliable means to track the topology with far-field techniques. Our results open an avenue for reconfigurable photonic devices capable of plasmon polariton steering, such as canaling, focusing and routing, and pave a way for low-symmetry plasmonic nanophotonics based on anisotropic natural materials.

**Introduction**

Hyperbolic polaritons are a unique type of polariton that exhibits hyperbolic isofrequency contours (IFCs). They are advantageous over traditional isotropic or elliptic polaritons. With extreme anisotropy, the propagation of hyperbolic polaritons is highly directional[1-5]. Meanwhile, they exhibit intense confinement and strong field enhancement, which enable sub-wavelength control of light-matter interactions, making them ideal candidates for applications such as sensing and energy conversion[6]. Additionally, the open geometry of the IFC in the momentum space leads to the theoretically infinite wavevectors and unprecedentedly high photonic density of states, which is particularly appealing in quantum applications like the enhancement of spontaneous emission[7].

Hyperbolic polaritons are typically found in man-made metamaterials, which require complicated nano-fabrication[8,9]. Fortunately, some anisotropic materials in nature have been discovered to host in-plane hyperbolic polaritons, such as phonon polaritons in $MoO_3$[10,11] and $V_2O_5$[12], and plasmon polaritons in $WTe_2$[13], which open up a plethora of opportunities in reconfigurable on-chip integrated photonics[14,15]. These findings have fueled an emerging research field termed low-symmetry nanophotonics[16-19]. Crucially, the tunability of the wavelength range of the hyperbolic IFCs (the hyperbolic regime) and the optical topological transition (OTT) of the IFCs to the elliptic topology[20] are highly desirable to fulfill the potential. However, phonon polaritons, which are based on polar lattice vibrations, intrinsically show rather fixed Reststrahlen bands and material-specific polariton dispersions. This certainly limits their tunability, although much efforts have been devoted to extrinsic schemes in $MoO_3$ phonon polaritons, such as interfacing with suitable substrate[21-24], stacking with highly tunable graphene[25-29], and twisting bilayer structures[30-33]. Progress has also been achieved through more disruptive measures, such as intercalation[12,34], but the phonon bands only shift a small amount[12] (around 30 cm$^{-1}$), and hyperbolic polariton was not observed after intercalation.

On the other hand, plasmon polaritons, particularly in the two-dimensional form, are easier to be tamed intrinsically[35,36], which has been exemplified by graphene plasmon polaritons[37]. However, naturally existing in-plane hyperbolic plasmon polaritons are rare, but have recently been demonstrated in WTe$_2$ thin films[13] in the frequency range of 429 cm$^{-1}$ to 632 cm$^{-1}$, with the elliptic regime of the IFCs below 429 cm$^{-1}$. Given the limited options for such materials, it's even more imperative to tune the OTT energy to suit various applications. In fact, the plasmon dispersion in a natural hyperbolic plasmonic surface is typically governed by the anisotropy of both free carrier response and bound interband transitions[13,38], with the former for inductive and the latter for capacitive optical responses. Any variation of electronic properties, such as carrier density, effective mass anisotropy, frequency and strength of interband transition resonance, will give rise to a modulation of the hyperbolic regime[38,39]. Though as promising as it sounds, however, the experimental demonstration of tunable in-plane OTT of IFCs of plasmon polaritons in a natural material has not been realized up to date.

In this study, we report the first intrinsic tuning of such OTT in a broad wavelength range through chemical doping and temperature in a natural material. We reveal such tunability in Mo-doped WTe$_2$ (Mo$_x$W$_{1-x}$Te$_2$) thin films, a recently discovered layered Type-II Weyl semimetal with composition-dependent band structure[40,41] and electric transport[42].

Meanwhile, an innovative technique to track the topology based on the far-field polarization dependence of the localized surface plasmon resonances (LSPRs) in skew ribbons has been developed. This technique allows for efficient and accurate characterization of the topology of plasmon dispersion (IFCs) at a particular frequency in a single sample. Our study not only extends the hyperbolic regime in natural materials by other degrees of freedom, but also reveals the peculiar and informative polarization property of LSPRs in microstructures made from an anisotropic material.

## Results

### Sample fabrication and polarized IR spectra

We grew $Mo_xW_{1-x}Te_2$ crystals ($x \leq 0.5$) in the semi-metallic orthorhombic $T_d$-phase using a chemical vapor transport technique with iodine as the transport agent (Materials and methods, Supplementary Note 1). The zigzag W-W chains are along $a$-axis, with W atoms partially substituted by Mo after Mo-doping[43], as displayed in Fig. 1a. Fig. 1b shows the schematic illustration of a skew ribbon array patterned from an exfoliated single crystal film of $Mo_xW_{1-x}Te_2$ with a skew angle of $\theta = -33°$ with respect to $a$-axis (Materials and methods), and illuminated by the normal incident light with a polarization angle $\phi$. LSPRs can be excited in such ribbon arrays with far-field incident light. Fig. 1c-d displays two typical extinction spectra (characterized by $1 - T/T_0$, where $T$ and $T_0$ are the transmission of light through the sample and the bare substrate, respectively) of such skew ribbon arrays with composition ratio $x = 0.278$ but different ribbon widths $L$ (corresponding to the effective plasmon wavevector of $\pi/L$). The light polarization ($\phi = $ -14.8°, 24.5°) was selected as close as possible to where the plasmon resonance is most intense, with resonance frequencies

(219 cm$^{-1}$, 462 cm$^{-1}$) in the elliptic and the hyperbolic regimes respectively (to be discussed below). Besides the plasmon resonance, the spectrum in Fig. 1d exhibits evident Drude response, in sharp contrast to that in Fig. 1c, suggesting that the polarization for maximal plasmon intensity in the hyperbolic regime deviates significantly from the perpendicular direction of ribbons. Such deviation was first reported in self-assembled carbon nanotubes[44]. However, the implication on the topology of plasmon dispersion has not been revealed. Here, we show that such optimal polarization is fully dictated by the ratio of the imaginary parts of the anisotropic conductivities, and in turn can be utilized to determine the topology of plasmon dispersion.

**Polarization dependence of LSPRs in skew ribbon arrays**

When a skew ribbon array is illuminated, the plasmon resonance is most intense when the polarization of the incident light $E_{\text{ext}}$ is parallel to the polarization current density $J_{\text{polar}}$ ($J_{\text{polar}} = \partial P/\partial t$, with $P$ as the polarization vector), which is associated with the depolarization field $E_{\text{depol}}$ induced by the polarization charge. Note that $E_{\text{depol}}$ is always perpendicular to the ribbon edge due to the translation-invariance of the polarization charge distribution along the edge (Fig. 1e-f), and $J_{\text{polar}}$ is not the conduction current related to the real part of the conductivity, which is responsible for the energy dissipation in the material. In an isotropic two-dimensional material, the incident light with polarization perpendicular to the ribbon leads to the maximal plasmon resonance, since $P$ and hence $J_{\text{polar}}$ are parallel to $E_{\text{depol}}$ due to the isotropic polarizability (conductivity) tensor. In ribbons made from an anisotropic film, however, $P$ and $J_{\text{polar}}$ are not necessarily parallel to $E_{\text{depol}}$. Thus, as shown in Fig. 1e-f, the optimal polarization $\phi_{\text{max}}$ for plasmon excitation deviates from the perpendicular direction of the ribbon, the value of which is determined by the ratio of the imaginary parts of conductivities ($\sigma''_{aa}$ and $\sigma''_{bb}$) and the skew angle $\theta$ (Supplementary Note 2):

$$\tan \phi_{\max}(\omega) = \frac{\sigma''_{bb}(\omega)}{\sigma''_{aa}(\omega)} \tan \theta \tag{1}$$

where $\phi_{\max}$ can be restricted in the range of -90° to 90°. Therefore, $\phi_{\max}$ has the same sign as $\theta$ in the elliptic regime since $\sigma''_{aa}\sigma''_{bb} > 0$ (Fig. 1e), but opposite sign in the hyperbolic regime, for which $\sigma''_{aa}\sigma''_{bb} < 0$ (Fig. 1f). Particularly, for the plasmon polaritons at the OTT energy ($\sigma''_{bb} = 0$), the optimal light polarization coincides with $a$-axis.

To benchmark this scheme, we firstly use Eq. 1 to reexamine the topology of plasmon dispersion in WTe$_2$ films, for which an OTT has been reported at about 429 cm$^{-1}$ (23.3 microns in wavelength)[13]. Skew ribbon arrays as in Fig. 1b with the same skew angle of $\theta = -33°$ but different ribbon widths were fabricated from WTe$_2$ films. It should be noted that all of the studied skew ribbons in this paper have the same -33° skew angle (Supplementary Note 3). The polarization-resolved extinction spectra for two representative samples at 10 K are shown in Fig. 2a-b (Materials and methods). Resonance peaks of LSPRs can be observed at frequencies of 308 cm$^{-1}$ in the elliptic regime (Fig. 2a) and 510 cm$^{-1}$ in the hyperbolic regime (Fig. 2b), thus expected to have maximal plasmon intensity at polarization angles of different signs, according to Eq. 1. Note that, in Fig. 2b, spectra with polarization at around -33° (perpendicular to the ribbon) are nearly flat, indicating almost no plasmon absorption, in striking contrast to ribbons patterned from isotropic films. To see the polarization dependence more clearly, the extinction spectra in Fig. 2a-b are plotted as pseudo color maps in Fig. 2c-d. The maximal plasmon intensity in Fig. 2c can be found below the zero line (red dashed line), while the plasmon resonance in Fig. 2d exhibits strongest absorption at an angle well above zero. The feature below 200 cm$^{-1}$ comes from the Drude response of free carriers, whose maximum is always along the ribbon ($\phi$ = 57°), a common scenario for both isotropic and anisotropic ribbons (Supplementary Note 4). To extract the plasmon weight, spectra were fitted with the Drude-Lorentz model (Materials and methods, Supplementary Note 5), and the fitted plasmon weight is plotted in Fig. 2e-f as a function of the polarization angle. As shown in Fig. 2e (Fig. 2f), the polar angle $\phi_{max}$ is -11.7° (35.1°), which has the same (opposite) sign as the skew angle (-33°), consistent with the elliptic (hyperbolic) topology of IFCs.

More systematically, a series of samples with the same skew angle ($\theta = -33°$) and incremental plasmon frequencies ranging from 185 cm$^{-1}$ to 510 cm$^{-1}$ were fabricated by varying the ribbon width. The polarization-resolved absorption spectra due to the plasmon resonance are plotted as pseudo color maps for several plasmon frequencies in Fig. 2g (Materials and methods). The angle for maximal plasmon intensity gradually evolves from negative to positive, crossing the zero line (*a*-axis) when the plasmon frequency coincides with the elliptic/hyperbolic boundary (429 cm$^{-1}$, determined by the plasmon dispersion in previous work[13]). The measured frequency-dependent polarization angle for the maximal plasmon intensity ($\phi_{max}$) is displayed in Fig. 3a as brown dots, with errors from both angle measurements and fittings (Supplementary Note 5). In addition, the simulations of the extinction spectra in WTe$_2$ skew ribbon arrays ($\theta = -33°$) with different ribbon widths were performed (Supplementary Note 6). The fitted results of the optimal $\phi_{max}$ are displayed as blue squares in Fig. 3a. The OTT of IFCs from the elliptic to the hyperbolic can be directly manifested by the sign change of the angle based on experiments and simulations, which agrees well with the calculations by Eq. 1 (black solid line) using the conductivities extracted from the plasmon dispersion in previous work[13] (Supplementary Note 7), validating the capability of tracking the OTT both qualitatively and quantitatively. Note that when the skew angle is fixed, $\phi_{max}$ is solely determined by $\frac{\sigma''_{bb}(\omega)}{\sigma''_{aa}(\omega)}$ at the plasmon frequency and will not be affected by the ribbon width or film thickness (assuming an unchanged band structure with sample thickness), enabling it to be a superior method to determine the topology of plasmon dispersion at individual frequencies.

**Mo-doping-dependent OTT of IFCs of plasmon polaritons**

With the convenient toolkit at our disposal, we can now proceed to investigate the tuning of such OTT. Ribbon arrays with the same configuration as shown in Fig. 1b were fabricated on Mo-doped $WTe_2$ films with various ribbon widths, resulting in different plasmon frequencies (Supplementary Note 8). The polarization dependence of plasmon polaritons was measured and the fitted $\phi_{max}$ values are summarized in Fig. 3b-c for 27.8% and 50% doping levels, respectively. A sign change for $\phi_{max}$ can be observed in Fig. 3b-c as the plasmon frequency increases, indicating an OTT in the doped samples as well. A large redshift of the zero-crossing point or the OTT frequency (indicated by the vertical black dashed line) occurs. It changes from 429 cm$^{-1}$ (23.3 microns) for $WTe_2$ to about 270 cm$^{-1}$ (37.0 microns) for $Mo_{0.5}W_{0.5}Te_2$. This corresponds to a 38% redshift in frequency or a 1.6-fold increase in wavelength for the OTT energy.

To further manifest the doping effect on the OTT, as an example, we compare the polarization dependence of plasmon polaritons with similar resonance frequencies of $345 \pm 15$ cm$^{-1}$ for different doping levels. As shown in Fig. 3d-f, the polarization angle for the maximal plasmon intensity increases from negative to positive upon doping. This can be seen even more clearly in Fig. 3g-i, which shows that the angle $\phi_{max}$ increases from –9.5° for WTe$_2$ films to 23.9° at 50% Mo-doping, suggesting the OTT of IFCs from the elliptic to the hyperbolic by Mo-doping. Note that the plasmon linewidth of these samples increases with doping (72 cm$^{-1}$ of WTe$_2$, 112 cm$^{-1}$ at 27.8% doping and 250 cm$^{-1}$ at 50% doping). The increasing linewidth is primarily due to the stronger coupling between plasmon polaritons and interband transitions, which have lower energy at larger doping levels and hence are closer to the plasmon mode, as shown in Fig. 4g. Such coupling-induced broadening is manifested in WTe$_2$ as well[13]. As shown in Fig. 2g, the plasmon linewidth increases as the frequency approaches that of the interband transition. The linewidth of intrinsic plasmon resonances at lower frequencies, hence largely free of interband coupling, is more similar for all Mo$_x$W$_{1-x}$Te$_2$ samples (with resonance frequencies of 295.0, 291.7, 214.3 cm$^{-1}$ and quality factors of 3.4, 3.1, 2.1 along $a$-axis at different doping levels respectively, Supplementary Note 9).

**Hyperbolic regime determined by the optical absorption in Mo$_x$W$_{1-x}$Te$_2$ thin films**

To delve into the physical mechanism of the tunable OTT, polarization-resolved extinction spectra of pristine films of Mo$_x$W$_{1-x}$Te$_2$ ($x$ = 0, 0.278, 0.5, with thicknesses of 110, 80, 42 nm respectively) were examined. The measured spectra (far- and mid-IR ranges) and their fitting curves (Supplementary Note 10) are plotted in Fig. 4a-c. The corresponding extracted sheet optical conductivity (imaginary part) is displayed in Fig. 4d-f, with the generally expanding hyperbolic regime marked by the shaded area based on their signs. The IFCs of plasmon dispersion in Mo$_x$W$_{1-x}$Te$_2$ are also plotted according to the loss function[13] to visualize the OTT (Supplementary Note 11). The lower boundary of the hyperbolic regime redshifts upon doping, consistent with the polarization behavior of LSPRs in skew ribbon arrays. In general, the topology of plasmon dispersion is determined by the optical response of both the intraband (Drude response) and interband transitions of carriers. The transition energy and oscillator strength of the interband transition resonance are determined by the specific band structure and the Fermi level, and the free carrier Drude response depends on the carrier density and effective mass. In our experiment, the extinction spectra in Fig. 4a-c share similar qualitative profiles. The primary difference in the far-IR spectra is quantitative in nature, such as the peak position of the first interband transition resonance, as shown in Fig. 4g. The transition energy decreases from 726 cm$^{-1}$ for WTe$_2$ to about 488 cm$^{-1}$ at 50% doping (Supplementary Note 10), as indicated by blue arrows for the spectra of *b*-axis polarization in Fig. 4a-c, which brings the whole interband transition feature to lower frequencies. This salient doping dependence governs the redshift of the hyperbolic lower boundary, as the dielectric (capacitive) part of the optical conductivity at long wavelength is primarily attributed to interband transitions. It is also worth noting that the fitted Drude weight (normalized with thickness) along *b*-axis slightly decreases upon Mo-doping (Supplementary Note 10), playing a minor role in the redshift of the OTT energy.

Furthermore, by substituting Im($\sigma$) in Fig. 4d-f into Eq. 1, the calculated frequency dependence of $\phi_{\mathrm{max}}$ is plotted as gray dashed lines in Fig. 3a-c, which are consistent with the directly measured angles (brown dots) in skew ribbon arrays. Hyperbolic boundary frequencies derived from the polarization dependence of skew ribbons (red dots), film extinction measurements (black right-pointing triangles) and the plasmon dispersion in Ref.

13 (blue pentagrams) are summarized in Fig. 4h. All these procedures show consistent doping dependence of the lower boundary, substantiating the redshift of the OTT energy.

**Temperature-dependent OTT of IFCs of plasmon polaritons**

As a typical semimetal, WTe$_2$ exhibits strong temperature dependence in its electronic properties[13,45], enabling active tuning of the OTT. To fulfill this potential, the polarization-resolved extinction spectra of a 35 nm thick WTe$_2$ bare film at different temperatures (78 K, 130 K, 155 K, 180 K, 230 K 300 K) were measured. The corresponding spectra (far- and mid-IR ranges) and fitting curves are plotted in Fig. 5a-d. The anisotropy of interband transitions (between 700 and 1100 cm$^{-1}$) along two axes decreases, whereas the Drude weights increase due to more thermal carriers at higher temperatures. As a result, the lower boundary of the hyperbolic regime blueshifts from 428 cm$^{-1}$ at 78 K to 553 cm$^{-1}$ at 180 K, which is manifested by the extracted sheet optical conductivity in Fig. 5e-h and the temperature dependence of IFCs (Supplementary Note 11). The hyperbolic regime in the far-IR range vanishes above 230 K. A kink appears in the temperature dependence of the hyperbolic regime at about 130 K (inset in Fig. 5l), which is consistent with the temperature dependence for the plasmon frequency and Drude weight in WTe$_2$[13]. This is likely attributed to the temperature-induced Lifshitz transition at about 147 - 160 K, where the two hole pockets move down in energy with respect to the Fermi surface and eventually disappear, resulting in no hole carriers at higher temperatures[46,47]. Further, the polarization dependence of plasmon polaritons of the same devices as those shown in Fig. 3a at corresponding temperatures was measured. The polarization angle $\phi_{\text{max}}$ was extracted in the same way, which agrees well with the calculation of Eq.1 as shown in Fig. 5i-l, demonstrating the temperature-induced shifts of the OTT energy. Additional details are summarized in Supplementary Note 12. Compared to nearly temperature-independent phonon polaritons, the hyperbolic regime in WTe$_2$ plasmon polaritons shifts over a wide range with temperature. By combining Mo doing and temperature tuning, the hyperbolic regime overall covers the far-IR spectrum from 13.5 microns (739 cm$^{-1}$) to 37.0 microns (270 cm$^{-1}$), which is 3.1 times broader than the hyperbolic wavelength range observed in pristine WTe$_2$ films at 10 K[13].

**Discussion**

**Characterization of the OTT in the far-IR range with the far-field method**

As a matter of fact, determining the topology of plasmon dispersion in the far-IR range is a daunting task without the aforementioned polarization-based method. In principle, both near- and far-field techniques can probe the OTT in WTe$_2$. However, the near-field scheme is not mature in the far-IR range, especially with samples in the cryogenic conditions, even though it is widely and successfully employed in the mid-IR range to image in-plane hyperbolic phonon polaritons[3-5,10-12,16,21-23,25-34]. As a consequence, up to now, there is no near-field imaging of hyperbolic plasmon polaritons in WTe$_2$. As for the far-field technique, previously we determined the OTT of plasmon polaritons in WTe$_2$ through mapping the plasmon dispersion in the whole two-dimensional momentum space[13], which was laborious and required numerous samples (each momentum $\boldsymbol{q}$ needs a ribbon array). Fortunately, with our polarization-based far-field method, we can now determine the topology of the IFC at a particular frequency in a single sample without invoking the whole plasmon dispersion, which is truly advantageous.

**Implications of the tuning of the OTT in WTe$_2$**

By leveraging Mo doping and temperature, the hyperbolic regime expands 3.1 times than that in pristine $WTe_2$. This significant broadening demonstrates that hyperbolic plasmon polaritons can be manipulated more readily, which is fundamentally different from previously reported hyperbolic phonon polaritons. For instance, the fabrication of $MoO_3/MoO_3$ twisted bilayers leads to a contraction of the hyperbolic regime[30-33,48]. As a result, the expanded hyperbolic regime now covers nearly the entire far-IR range. This expansion complements the existing hyperbolic phonon polaritons that predominantly reside in the mid-IR range. Moreover, the far-IR range contains a multitude of intramolecular or intermolecular vibration modes (e.g., in proteins or DNA), making $Mo_xW_{1-x}Te_2$ an excellent candidate for bio-sensing and bio-imaging applications[49]. Furthermore, both the lower and higher boundaries of the hyperbolic regime, which correspond to the sigma-near-zero points along *b*- and *a*-axis, respectively, experience substantial shifts. When a material exhibits near-zero effective permittivity (conductivity), novel physical effects arise, such as field enhancement, tunneling through anomalous waveguides and transmission with small phase variation, which are also known as epsilon-near-zero photonics[50]. Thus, $Mo_xW_{1-x}Te_2$ naturally serves as an in-plane tunable anisotropic sigma-near-zero material for functional photonic devices.

Particularly, at the lower hyperbolic boundary, the IFC comprises two nearly parallel lines along *b*-axis in the two-dimensional momentum space, analogous to the so-called canalization regime in hyperbolic phonon polaritons[3,30,32]. Thus, the tunability of the lower boundary allows for the canalization of the energy flow of hyperbolic polaritons over a wide spectral range. The estimated propagation length of canalized plasmon polaritons in $WTe_2$ reaches 0.5 micron[13], which is comparable to that of isotropic plasmon polaritons in graphene[51] and that of the canalized phonon polaritons in h-BN metasurfaces[3]. Additionally, the doping procedure does not degrade the sample quality much, as suggested by the similar figure of merits (Supplementary Note 9). The lifetime of plasmon polaritons in $Mo_xW_{1-x}Te_2$ is approximately 0.05 picosecond (ps), comparable to that of plasmon polaritons in the undoped $WTe_2$ (0.1 ps)[13] and in graphene on $SiO_2$/Si substrates (about 0.05 – 0.1 ps)[52], though it is smaller than the lifetime of phonon polaritons in $MoO_3$ (about 8 ps)[10]. Future efforts can be devoted to increasing the sample quality through more meticulous growth and judicious choice of substrates.

In conclusion, our work demonstrates the inherent tunability of hyperbolic plasmon polaritons and the OTT in vdW surfaces by chemical doping and temperature over a wide range. The tuning mechanism involves both bound states and free carriers, providing more dimensions for manipulating OTT. Our experiments leverage a unique feature in the polarization-resolved extinction spectra of skew ribbons to determine the topology of IFCs, which can be of great use to investigate other anisotropic two-dimensional materials.

**Materials and methods**

**Mo$_x$W$_{1-x}$Te$_2$ crystal growth**

Mo$_x$W$_{1-x}$Te$_2$ single crystals were grown by a chemical vapor transport technique with iodine as the transport agent. Stoichiometric mixtures of Mo, W and Te powders were loaded into a quartz tube along with a small amount of iodine, which was subsequently sealed in vacuum and placed in a two-zone furnace. The hot zone was maintained at 850 ℃ for two weeks while the cold zone was kept at 750 ℃. The composition of the final crystal was characterized using energy dispersive spectroscopy (EDS) with a scanning electron microscope.

**Sample preparation and fabrication**

Single crystal WTe$_2$ was bought from HQ Graphene. Bare films of Mo$_x$W$_{1-x}$Te$_2$ ($x \leq 0.5$) with thickness about 40 - 120 nm through the standard exfoliation technique were transferred to polycrystalline diamond substrates. The substrate has no polar phonon absorption and the transmission is about 70% in the far-IR range. The thickness of the sample was determined by a stylus profiler (Bruker DektakXT) in conjunction with the optical contrast. Preferred sample size is 200 by 200 µm$^2$, with the side length longer than the far-IR wavelength (~100 µm). Skew ribbon arrays were patterned using electron beam lithography (Zeiss Sigma SEM with Raith Elphy Plus), with the uncertainty in skew angle of less than 0.5°. An intermediate skew angle of −33° was chosen in our study, because a too small angle results in a small $\phi_{\max}$ as shown in Eq. 1, and a too large one leads to a more dramatic reduction of the maximal plasmon frequency achievable in that wavevector direction. Reactive ion etching (RIE) with SF$_6$ gas was used to define the ribbons. If necessary, the surroundings of the sample were further etched away using the director writer UPG501 and RIE to ensure IR response only from the targeted sample.

**Far-IR optical spectroscopy**

For the polarized far-IR extinction spectra, we used a Bruker FTIR spectrometer (Vertex 70 v) integrated with a Hyperion 2000 microscope and a cryogen-free silicon bolometer system as the detector. The incident light was focused on $Mo_xW_{1-x}Te_2$ samples with a 15× IR objective. A THz polarizer was used to control the light polarization. The samples were cooled to 10 K in a helium-flow cryostat (Janis Research ST-300) with vacuum at about $5 \times 10^{-5}$ mbar. Throughout the entire measurements, compressed dry air with dew point below -70 °C was purged to an enclosed space housing the cryostat. This procedure minimized the absorption of IR light by moisture in air and effectively increased the signal/noise ratio. The polarization dependence of plasmon polaritons in skew ribbon arrays was studied by rotating the polarizer with a step size of 11.3° from -90° to 90°. Hence, a total of 16 spectra were collected to extract each $\phi_{max}$.

**Fitting of IR extinction spectra and plotting pseudo color maps for plasmon spectra**

The extinction spectrum is determined by the sheet optical conductivity $\sigma(\omega)$ as follows[52]:

$$1 - \frac{T}{T_0} = 1 - \frac{1}{|1+Z_0\sigma(\omega)/(1+n_s)|^2} \qquad (2)$$

where $Z_0$ is the vacuum impedance, $\omega$ is the frequency of light, and $n_s$ is the refractive index of the substrate. The conductivity of the sample is expressed by the Drude-Lorentz model, where the Drude model describes free carriers and the Lorentz model accounts for the bound states such as plasmon resonance or interband transitions in our system:

$$\sigma(\omega) = \frac{i}{\pi}\frac{D}{\omega+i\Gamma} + \sum_k \frac{i}{\pi}\frac{\omega S_k}{\omega^2-\omega_k^2+i\omega\Gamma_k} \qquad (3)$$

In Eq. 3, $D$ and $S_k$ represent the spectrum weights, $\omega_k$ represents the resonance frequency of the plasmon resonance or interband transitions, $\Gamma$ and $\Gamma_k$ are the corresponding FWHMs. The polarization angle $\phi_{max}$ was extracted by fitting the whole set of 16 polarization spectra with one Drude component and two Lorentz components (plasmon resonance, interband transitions respectively). The spectral weights were fitting parameters and the Drude scattering rate, plasmon frequency and interband transition resonance frequency were kept the same for all spectra. Pseudo color maps in Fig. 2g and Fig. 3d-f were plotted by substituting the fitted conductivities of plasmon resonances into Eq. 2 in turn. The plasmon weights $S_p$ extracted above were fitted as $\cos^2\phi$ to obtain $\phi_{max}$. The fitting details of the extinction spectra of $Mo_xW_{1-x}Te_2$ bare films in Fig. 4a-c are discussed in Supplementary Note 10.

**Acknowledgments**

H.Y. is grateful to the financial support from the National Key Research and Development Program of China (Grant Nos. 2022YFA1404700 and 2021YFA1400100), the National Natural Science Foundation of China (Grant No. 12074085), the Natural Science Foundation of Shanghai (Grant No. 23XD1400200). C.W. is grateful to the financial support from the National Natural Science Foundation of China (Grant Nos. 12274030, 11704075) and the National Key Research and Development Program of China (Grant No. 2022YFA1403400). F.S. acknowledges the financial support from the National Key Research and Development Program of China (Grant No. 2017YFA0303203), the National Natural Science Foundation of China (Grant Nos. 92161201, 12025404, 11904165, and 12274208), the Natural Science Foundation of Jiangsu Province (Grant No. BK20190286). S.H. is grateful to the financial support from the China Postdoctoral Science Foundation (Grant No. 2020TQ0078). Part of the experimental work was carried out in Fudan Nanofabrication Lab.

**Conflict of interests**

The authors declare no competing financial interests.

**Contributions**

H.Y. and C.W. initiated the project and conceived the experiments. Y.X. prepared the samples, performed the measurements and data analysis with assistance from C.W., Y.D., Yuqi Li, Q.X., S.H., Yuchen Lei, L.M. and J.Z. F.F. and F.S. grew and characterized the bulk single crystals of $Mo_xW_{1-x}Te_2$. H.Y. and C.W. and Y.X. co-wrote the manuscript. H.Y. supervised the whole project. All authors commented on the manuscript.

**Data availability**

All data needed to evaluate the conclusions in the paper are present in the main text and the supplementary information.

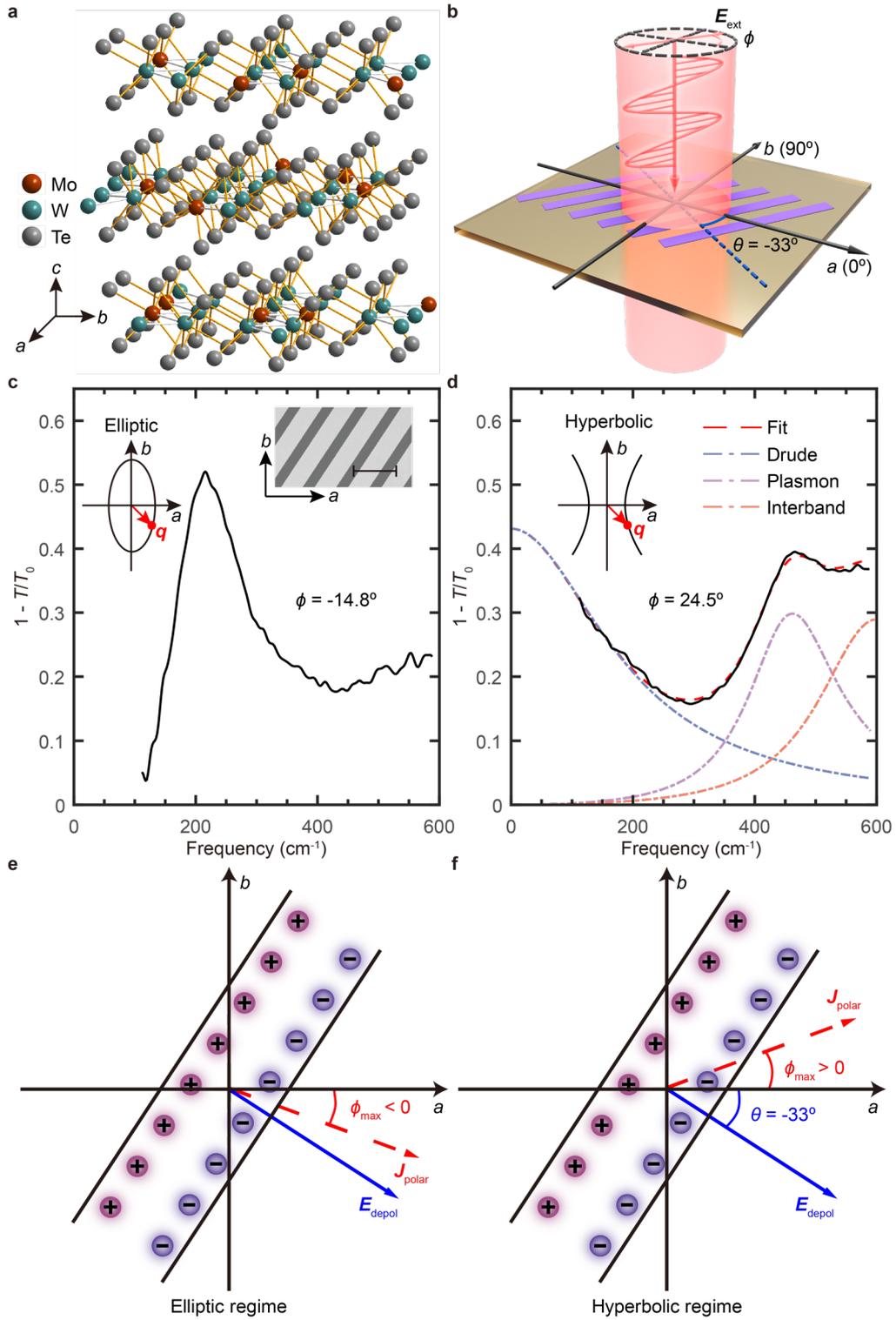

**Fig. 1 Polarization $\phi_{max}$ for the maximal plasmon intensity. a** A sketch of Mo-doped WTe$_2$ in $T_d$-phase. Brown/green represents Mo/W atoms, gray represents Te atoms. **b** Schematic illustration of skew ribbon arrays under illumination. Counter clockwise rotation with respect to $a$-axis is defined as positive. The polarization of normal incident light is $\phi$

and the perpendicular direction of the ribbon is $\theta = -33°$. **c, d** The typical extinction spectra of skew ribbon arrays (composition ratio $x = 0.278$, ribbon widths of 4.5 and 1.9 µm) along the polarization ($\phi = -14.8°, 24.5°$) in the elliptic/hyperbolic regime respectively. Right inset in **c** shows the SEM image of the skew ribbon array. Scale bar is 10 µm. Top left insets in **c** and **d** are schematics of the IFCs of plasmon mode in the elliptic and hyperbolic regimes respectively. Red arrows labelled by **q** are the corresponding wavevector direction (related to the $-33°$ skew angle). **e, f** The configuration of polarization current density $J_{polar}$ and depolarization field $E_{depol}$ in elliptic and hyperbolic regimes respectively. $\phi_{max}$ has the same sign as $\theta$ in the elliptic regime, but opposite sign in the hyperbolic regime.

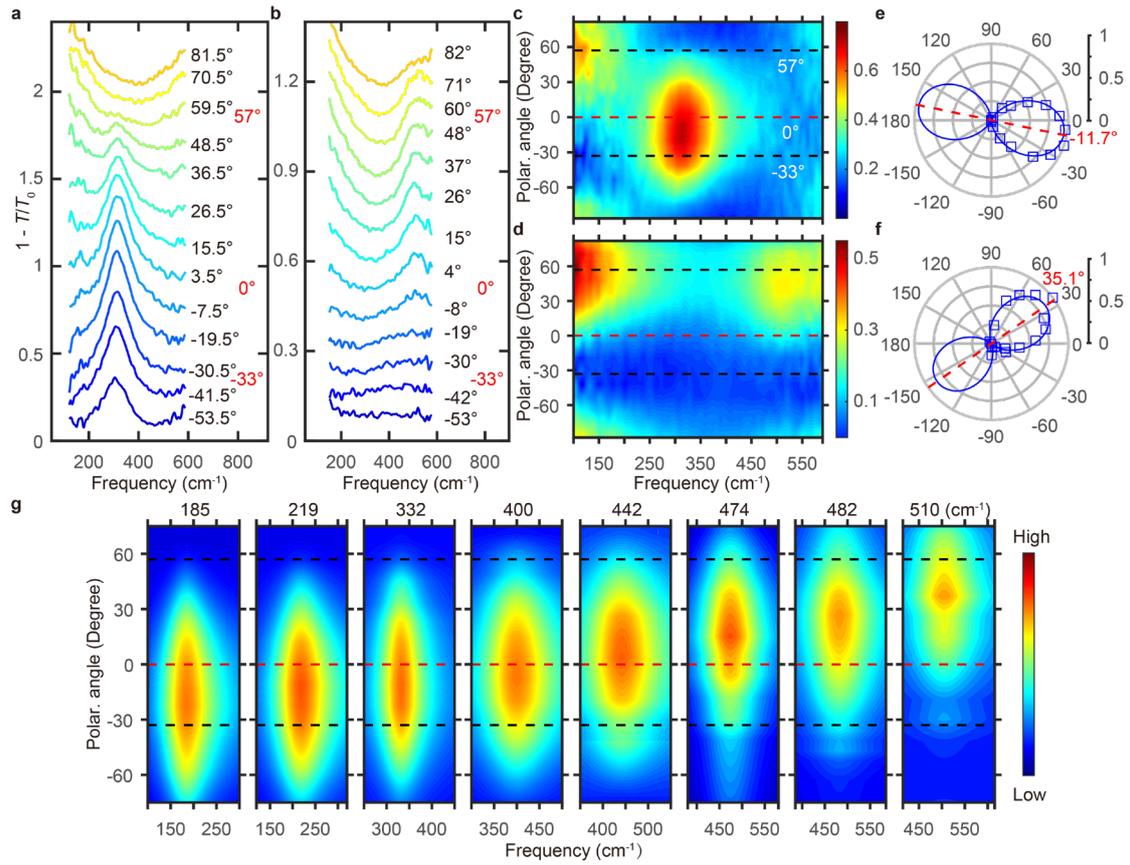

**Fig. 2 Polarization-resolved LSPRs in skew ribbon arrays of WTe$_2$. a, b** Polarization dependence of raw extinction spectra at plasmon frequencies of 308 cm$^{-1}$ and 510 cm$^{-1}$ respectively. Degrees in black indicate the polarization of incident light. Angles of 57° and −33° in red represent the parallel and perpendicular directions of ribbon edges. 0° denotes *a*-axis of WTe$_2$. For clarity, spectra are shifted vertically. **c, d** The corresponding pseudo color maps of **a** and **b**. **e, f** Polarization dependence of the corresponding normalized plasmon weights of **a** and **b**. Blue solid lines are fitting results of $\cos^2\phi$. **g** The pseudo color maps of the plasmon absorption spectra at different resonance frequencies. In **c**, **d** and **g**, black dashed lines denote the parallel (57°) and perpendicular (−33°) directions with respect to the ribbon and the red dashed lines denote *a*-axis.

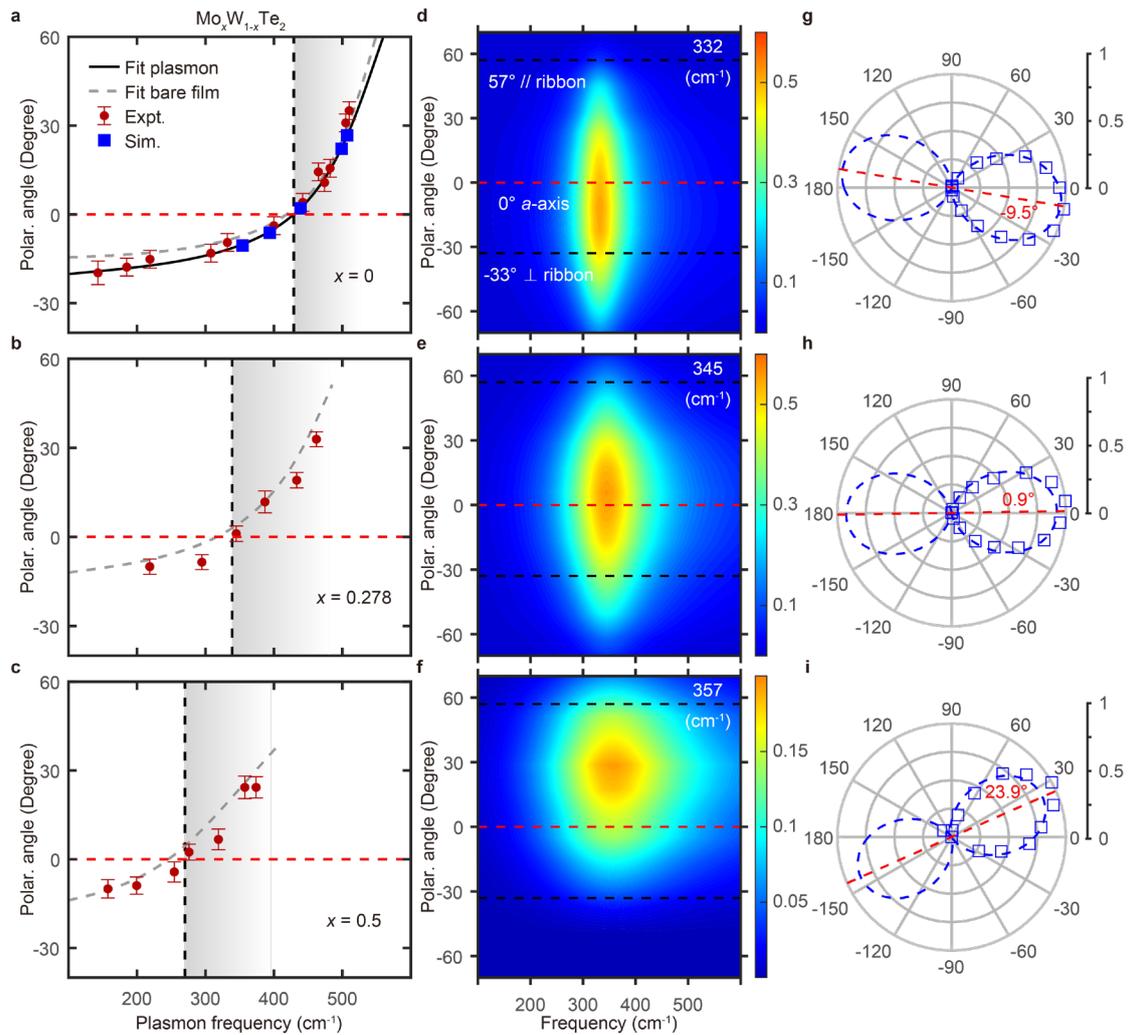

**Fig. 3 Tunable OTTs by Mo-doping. a-c** The polarization $\phi_{max}$ as a function of the plasmon frequency at different doping levels. Brown dots are fitted results of polarization experiments with errors of about 3° from both angle measurements and fittings. Blue squares are fitted results of simulations of the same configurations (Supplementary Note 6). Vertical black dashed line highlights the lower boundary of the hyperbolic regime. Black solid curve in **a** is based on the conductivity from the plasmon dispersion in Ref. 13. Gray dashed lines in **a**, **b** and **c** are calculated using the fitted conductivities from the extinction spectra of unpatterned films in Fig. 4a-c. **d-f** The pseudo color maps of the plasmon spectra with similar resonance frequencies (345 ± 15 cm$^{-1}$) at different Mo-doping levels (0, 0.278, 0.5 from top to bottom). **g-i** The corresponding normalized plasmon weights of **d-f** versus the polarization.

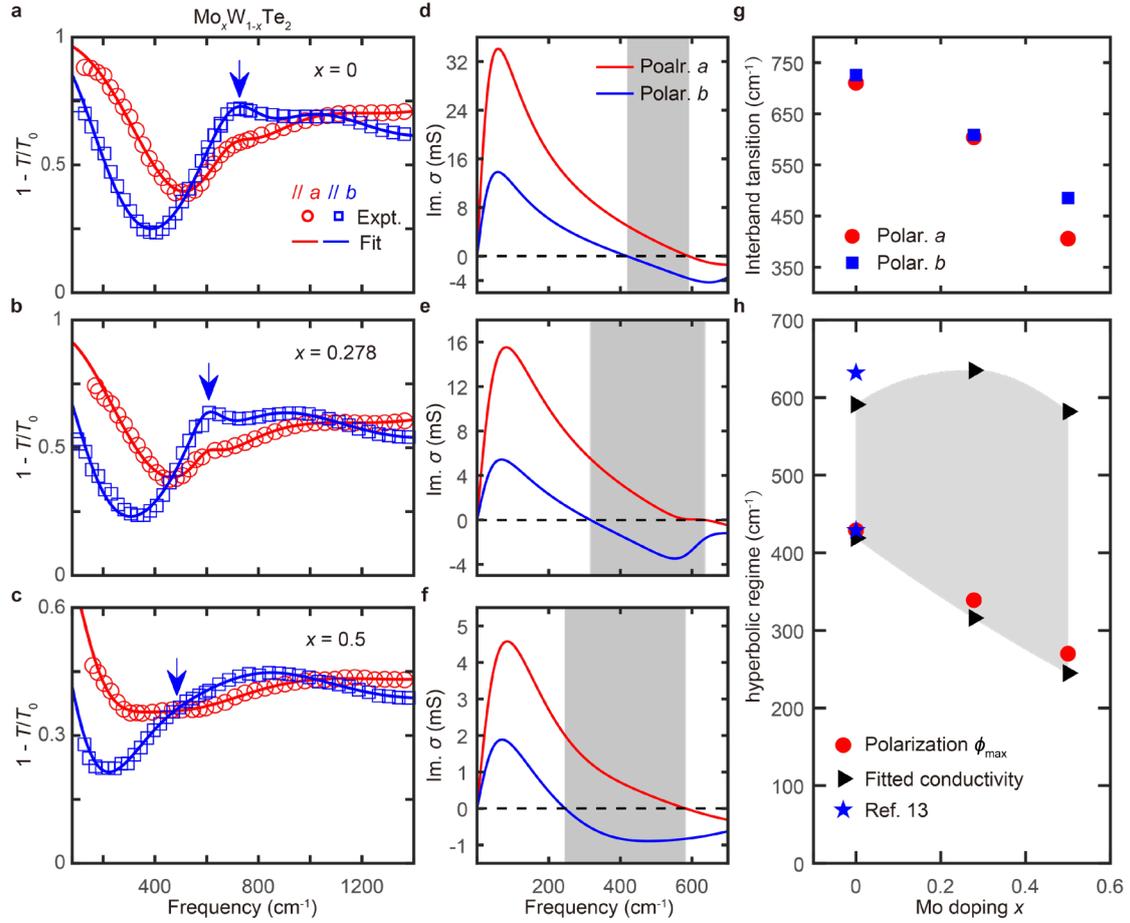

**Fig. 4 The mechanism of tunable OTTs by Mo doping. a-c** The experimental and fitting results of anisotropic extinction spectra of bare films of $Mo_xW_{1-x}Te_2$. Blue downward arrows represent the first interband transition resonances along *b*-axis. **d-f** The imaginary parts of conductivities along two crystal axes. Shaded areas denote the corresponding hyperbolic regimes. **g** Doping dependence of the frequency of the first interband transition resonance. **h** Doping dependence of the hyperbolic regime. Black right-pointing triangles are determined by fitted conductivities from **d-f** and shaded areas are the corresponding hyperbolic regimes. Red dots are determined by the polarization $\phi_{max} = 0$. Blue pentagrams are from the plasmon dispersion in Ref. 13.

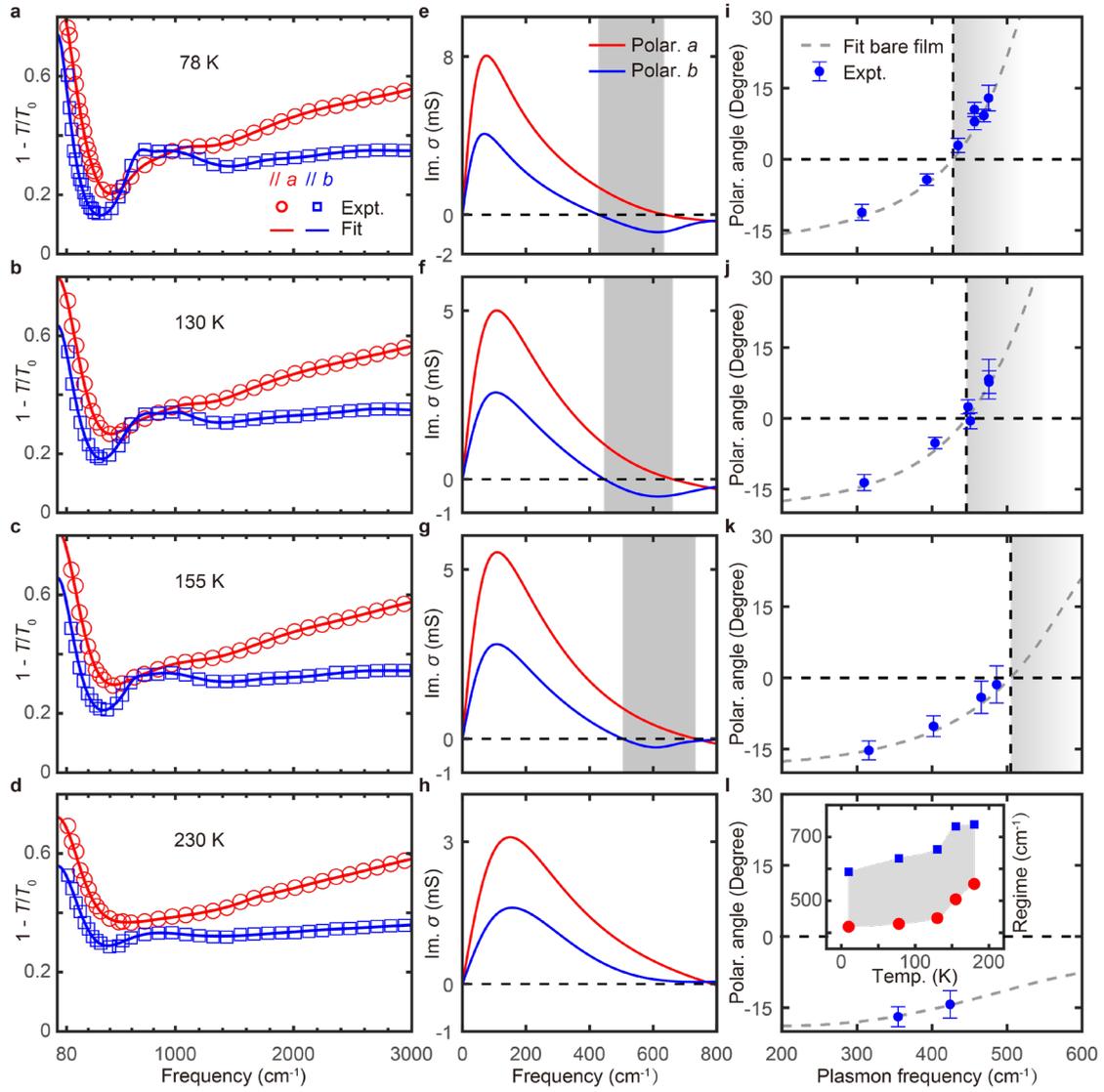

**Fig. 5 Temperature-induced shifts of OTT energy in WTe$_2$. a-d** The anisotropic extinction spectra and fitting curves of a bare film of WTe$_2$ at 78, 130, 155, and 230 K respectively. **e-h** The corresponding extracted imaginary parts of the optical conductivities along two crystal axes in **a-d**. Shaded area represents the hyperbolic regime. **i-l** The optimal polarization $\phi_{\text{max}}$ as a function of the plasmon frequency at 78, 130, 155, and 230 K respectively. Dashed lines are calculated using the fitted optical conductivities in **e-h** at the corresponding temperature. Inset in **l** shows the temperature dependence of the hyperbolic regime determined by the corresponding extracted imaginary part of the optical conductivity (Supplementary Note 12).

# Supplementary Information for
# Tunable optical topological transitions of plasmon polaritons in WTe$_2$ van der Waals films

Yuangang Xie[1], Chong Wang[2,3]*, Fucong Fei[4,5]*, Yuqi Li[2,3], Qiaoxia Xing[1], Shenyang Huang[1], Yuchen Lei[1], Jiasheng Zhang[1], Lei Mu[1], Yaomin Dai[6], Fengqi Song[4,5], Hugen Yan[1]*

*Corresponding author. Email: chongwang@bit.edu.cn (C.W.), feifucong@nju.edu.cn (F.F.), and hgyan@fudan.edu.cn (H.Y.).

**Supplementary notes:**

1: Characterization of the Composition of $Mo_xW_{1-x}Te_2$.

2: Polarization Dependence of LSPRs in Skew Ribbon Arrays.

3: The Selection of the Skew Angle *θ*.

4: Polarization Dependence of the Drude Response in Skew Ribbon Arrays.

5: Fitting Details and Errors of LSPRs in Skew Ribbon Arrays.

6: Simulations of Polarized Extinction Spectra in Skew Ribbon Arrays.

7: Conductivities Extracted from the Plasmon Dispersion of $WTe_2$.

8: Ribbon Width Dependence of LSPRs in $Mo_xW_{1-x}Te_2$ Skew Ribbons.

9: LSPRs in Relatively Large $Mo_xW_{1-x}Te_2$ Disks.

10: Fitting of the Extinction Spectra of Bare Films and the Extracted Drude Weights.

11: IFCs of Plasmon Dispersion in $Mo_xW_{1-x}Te_2$.

12: Temperature Dependence of the OTT in $WTe_2$.

# 1. Characterization of the Composition of $Mo_xW_{1-x}Te_2$.

The final crystal composition was characterized using energy dispersive spectroscopy (EDS) with a scanning electron microscope. EDS measurements were taken at multiple locations on the crystal surface to obtain an average composition. Fig. S1a and b show the typical EDS at 27.8% and 50% Mo doping, respectively.

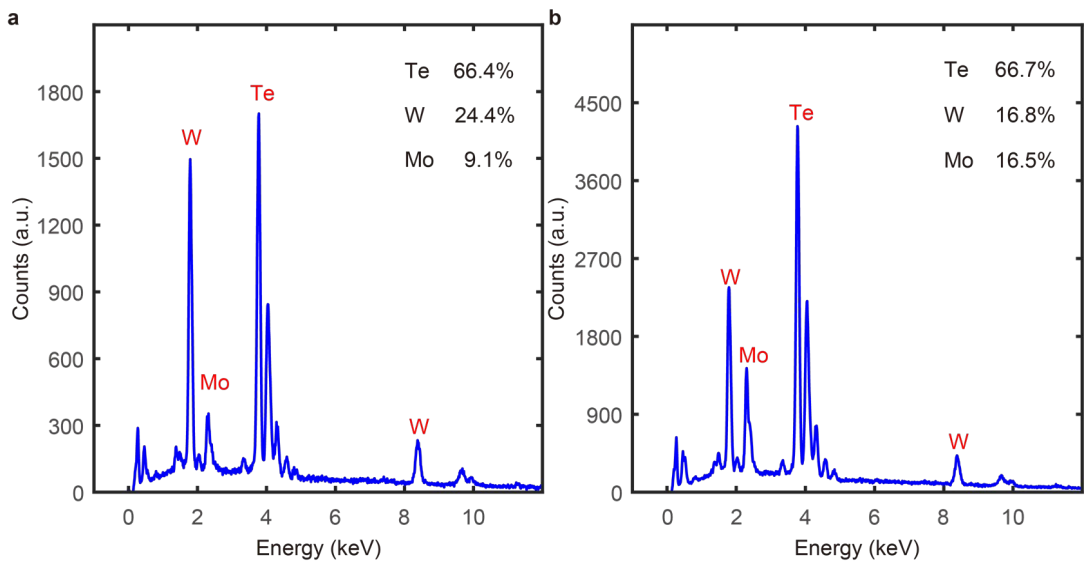

**Fig. S1 | The typical EDS at 27.8% (a) and 50% (b) Mo doping.**

# 2. Polarization Dependence of LSPRs in Skew Ribbon Arrays.

Equation (1) in the main text fully captures the polarization dependence and we derive it in detail now. Here the Ohmic loss (real part of the conductivity) is neglected for convenience. The conclusion, however, is general and rigorous. LSPRs can be treated as self-sustained dipoles, mimicking mechanical oscillators, which give rise to the polarization current density $J_{polar}$ and the depolarization field $E_{depol}$. Note that the LSPR in the ribbons is independent of the existence of the external field, which only serves as the external stimuli to start the plasmon dipole oscillation and to compensate the damping afterwards, if there is any. Again, this is fully analogous to a mechanical damped oscillator. $J_{\text{polar}} \cdot E_{\text{ext}}$ accounts for the energy provided by the external field to the plasmon system. Therefore, the most efficient way to drive and feed the plasmon resonance is to let $E_{\text{ext}}$ trace the polarization current density $J_{\text{polar}}$, that is, $E_{\text{ext}}$ is parallel to $J_{\text{polar}}$, as displayed in Fig. S2.

The polarization current density $J_{\text{polar}}$ is equal to $\frac{\partial P}{\partial t}$. Here, $P$ is the electric polarization caused by the depolarization field (rather than the external field) $P = \overleftrightarrow{\chi}\varepsilon_0 E_{\text{depol}} \propto \overleftrightarrow{\sigma''} E_{\text{depol}}$, where $\overleftrightarrow{\chi}$ is the polarizability tensor. The polarization $P_{\text{ext}}$ directly generated by the external field $E_{\text{ext}}$ is in-phase with $E_{\text{ext}}$, hence the associated current $\frac{\partial P_{\text{ext}}}{\partial t}$ is out of phase and doesn't take energy from the external field. Note that $E_{\text{depol}}$ is always perpendicular to the ribbon since polarization charges are along the two edges of the long ribbon, so:

$$J_{\text{polar}} \propto \begin{pmatrix} \sigma''_{aa} & 0 \\ 0 & \sigma''_{bb} \end{pmatrix} \cdot \begin{pmatrix} |E_{\text{depol}}| \cos\theta \\ |E_{\text{depol}}| \sin\theta \end{pmatrix} \propto \begin{pmatrix} \sigma''_{aa}(\omega) \cos\theta \\ \sigma''_{bb}(\omega) \sin\theta \end{pmatrix} \tag{S1}$$

where $\sigma''_{aa}$ and $\sigma''_{bb}$ are the imaginary parts of the diagonal elements of the anisotropic conductivity tensor, and $\theta$ is the skew angle of the ribbon array. The polarization of the incident light for the maximal plasmon intensity is defined as $\phi_{\text{max}}$ in the main text. Thus, at such polarization, we have $E_{\text{ext}} = \begin{pmatrix} |E_{\text{ext}}| \cos\phi_{\text{max}} \\ |E_{\text{ext}}| \sin\phi_{\text{max}} \end{pmatrix}$, which should be parallel to $J_{\text{polar}}$ as discussed above. In conjunction with equation (S1), we arrive at:

$$\tan\phi_{\text{max}}(\omega) = \frac{\sigma''_{bb}(\omega)}{\sigma''_{aa}(\omega)} \tan\theta \tag{S2}$$

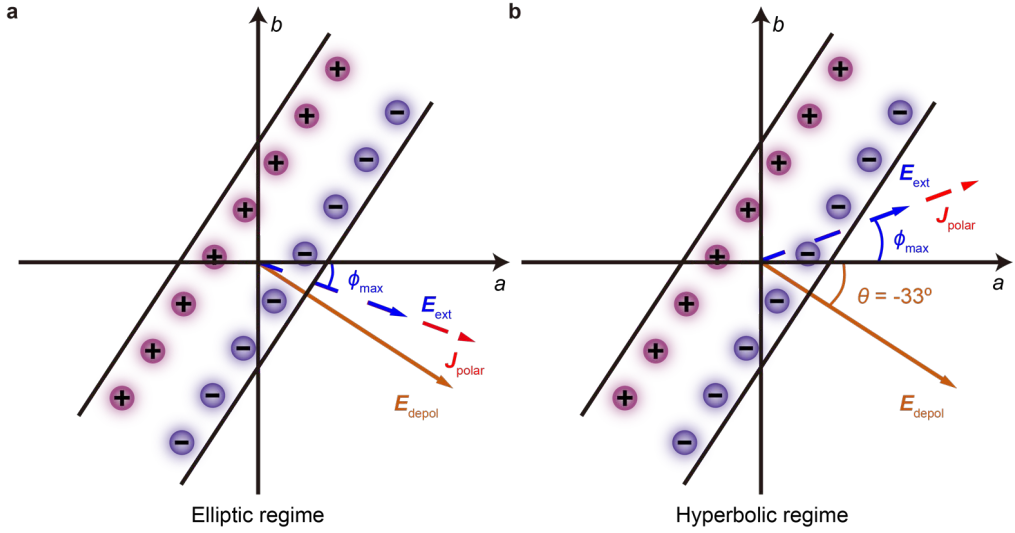

**Fig. S2 | Polarization dependence of plasmon spectra in skew ribbon arrays in the elliptic and hyperbolic regimes respectively.** The optimal polarization for $\boldsymbol{E}_{\text{ext}}$ is indicated.

### 3. The Selection of the Skew Angle $\theta$.

A skew angle of -33° is chosen based on the following factors: 1), the value of the optimal polarization $\phi_{\max}$ at each resonance frequency $\omega$; 2), the maximal resonance frequency $\omega_{\max}$ we can achieve in the skew ribbon arrays with the skew angle $\theta$. Note that we are discussing the absolute value here because the sign of the skew angle (or the polarization angle $\phi$) depends on our definition arbitrarily and does not affect the conclusion.

Firstly, according to equation (1) or (S2), $\phi_{\max}(\omega)$ increases with $\theta$, as shown in Fig. S3. Thus, a large $\theta$ is needed to make $\phi_{\max}(\omega)$ large enough for detection in experiments.

Secondly, $\omega_{\max}$ decreases with the increase of $\theta$. As we know, the ribbon configuration directly determines the wavevector of the plasmon polariton[1]. Specifically, for the ribbon with width $L$, the effective wavevector $\boldsymbol{q}$ is $\pi/L$, and the direction of $\boldsymbol{q}$ is perpendicular to the ribbon. In the elliptic regime, a smaller ribbon width gives a higher plasmon frequency, as the resonance frequency of plasmon polariton typically increases with the wavevector (e.g., the $\omega \propto \sqrt{q}$ relation in the long-wavelength limit). However, in the hyperbolic regime, the dispersion is softened due to the coupling with interband transitions. For example, the maximal resonance frequencies of WTe$_2$ along $a$- and $b$-axis are about 632 and 429 cm$^{-1}$, respectively[2]. When the wavevector direction rotates from $a$- to $b$-axis (i.e., when $\theta$ increases), the maximal plasmon frequency decreases. Such effect has been confirmed by our previous work[2] (Supplementary Figure 9 in Ref. 2). As a result, a smaller $\theta$ can expand the range of the plasmon resonance frequency we can achieve in the skew ribbon arrays.

At last, a moderate skew angle of -33° was chosen in our study to ensure a relatively large $\phi_{max}(\omega)$ and $\omega_{max}$ simultaneously.

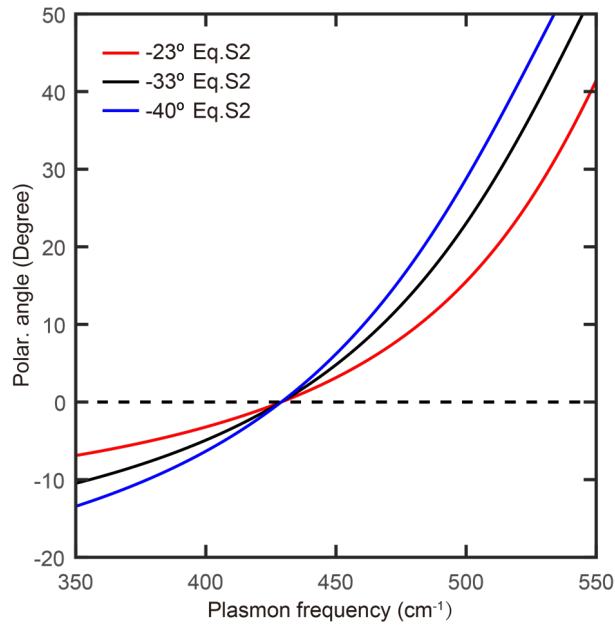

**Fig. S3 | The polarization $\phi_{max}$ as a function of the plasmon frequency at three skew angles (-40°, -33°, -23°).**

## 4. Polarization Dependence of the Drude Response in Skew Ribbon Arrays.

Free carriers are responsible for the Drude response with peak at zero frequency. Fig. S4a exhibits the polarization dependence of the normalized Drude weight of Fig. 2b in the main text. The Drude weight is largest when the polarization is along 57°, which is precisely parallel to the ribbon (skew angle is -33°). Interestingly, this is the same as that in ribbons made from isotropic two-dimensional films, such as graphene. Here, we'll give a justification for such a scenario. For simplicity and without loss of generality, the Drude response at zero frequency or the DC response is examined. The light absorption is related to the Joule heating, namely $\bm{J} \cdot \bm{E}$, where $\bm{J}$ and $\bm{E}$ ($\bm{E} = \bm{E}_{\text{ext}} + \bm{E}_{\text{depol}}$) are the total current density and electric field respectively. The geometrical shape of the ribbon results in the following two features: 1) the depolarization field $\bm{E}_{\text{depol}}$ is perpendicular to the ribbon due to the translation-invariance of the induced charge distribution at the edges, and 2) the current density component perpendicular to the ribbon edge $J_\perp$ must vanish, and only $J_\parallel$ survives after it reaches quasi-static state.

We can reach such a conclusion that the polarization for maximal Drude weight is always along the ribbon direction in two ways. The first way is to consider the situation when the external field is perpendicular to the ribbon. As we'll show, there is no current (either $J_\perp$ or $J_\parallel$) at all when it reaches the static state and no power is consumed. According to feature 1), the total electric field $\bm{E}$ is perpendicular to the ribbon as well. Therefore, to meet feature 2), $\bm{E}$ must be zero to get the zero value of $J_\perp$, i.e., the external field and depolarization field compensate each other (equal and opposite). As a result, there is no steady current when the external field is perpendicular to the ribbon, namely $\bm{J} = 0$ and no Joule heating (Drude response) at all. Since we have found the minimum, the maximal Drude response occurs at the polarization perpendicular to the minimal case, i.e., parallel to the ribbon.

The second way is to directly find the maximum for $\boldsymbol{J} \cdot \boldsymbol{E}$ at different polarization with constant external field amplitude $|\boldsymbol{E}_{\text{ext}}|$. As mentioned above (feature 1), $\boldsymbol{J}$ has a definitive direction which is parallel to the ribbon. According to the Ohm's law $\boldsymbol{J} = \overleftrightarrow{\sigma}\boldsymbol{E}$, where $\overleftrightarrow{\sigma}$ is the conductivity tensor,

$$\boldsymbol{J} = \begin{pmatrix} |\boldsymbol{J}| \cos \alpha \\ |\boldsymbol{J}| \sin \alpha \end{pmatrix} = \overleftrightarrow{\sigma}\boldsymbol{E} = \begin{pmatrix} \sigma_{aa} & 0 \\ 0 & \sigma_{bb} \end{pmatrix} \begin{pmatrix} |\boldsymbol{E}| \cos \beta \\ |\boldsymbol{E}| \sin \beta \end{pmatrix} = \begin{pmatrix} |\boldsymbol{E}|\sigma_{aa} \cos \beta \\ |\boldsymbol{E}|\sigma_{bb} \sin \beta \end{pmatrix} \quad (S3)$$

where $\sigma_{aa}$ and $\sigma_{bb}$ are the diagonal elements of the conductivity tensor (real parts), $\alpha$ and $\beta$ are the tilted angles of the ribbon and the total electric field $\boldsymbol{E}$ respectively, as shown in Fig. S4b. Therefore, for a given conductivity tensor, the total electric field $\boldsymbol{E}$ has a fixed orientation $\beta$:

$$\tan \beta = \frac{\sigma_{aa}}{\sigma_{bb}} \tan \alpha \quad (S4)$$

Now the Drude response is:

$$\boldsymbol{J} \cdot \boldsymbol{E} = (|\boldsymbol{E}|\sigma_{aa} \cos \beta \quad |\boldsymbol{E}|\sigma_{bb} \sin \beta) \cdot \begin{pmatrix} |\boldsymbol{E}| \cos \beta \\ |\boldsymbol{E}| \sin \beta \end{pmatrix} = |\boldsymbol{E}|^2 (\sigma_{aa} \cos \beta^2 + \sigma_{bb} \sin \beta^2) \quad (S5)$$

Obviously, the maximum of $|\boldsymbol{E}|$ leads to the maximal absorption, since $\beta$ is fixed according to equation (S4). Because $\boldsymbol{E}$ has a fixed direction and one of its components $\boldsymbol{E}_{\text{depol}}$ is always perpendicular to the ribbon, we can find its maximum geometrically. We draw a circle with radius $|\boldsymbol{E}_{\text{ext}}|$ (blue dashed circle), as shown in Fig. S4b. According to the vector summation rule (triangle rule) for $\boldsymbol{E} = \boldsymbol{E}_{\text{ext}} + \boldsymbol{E}_{\text{depol}}$, the external field $\boldsymbol{E}_{\text{ext}}$ parallel to the ribbon gives the maximal $|\boldsymbol{E}|$. Therefore, the Drude response is largest when the external field is applied along the ribbon direction. This rationalizes the experimental finding in Fig. S4a.

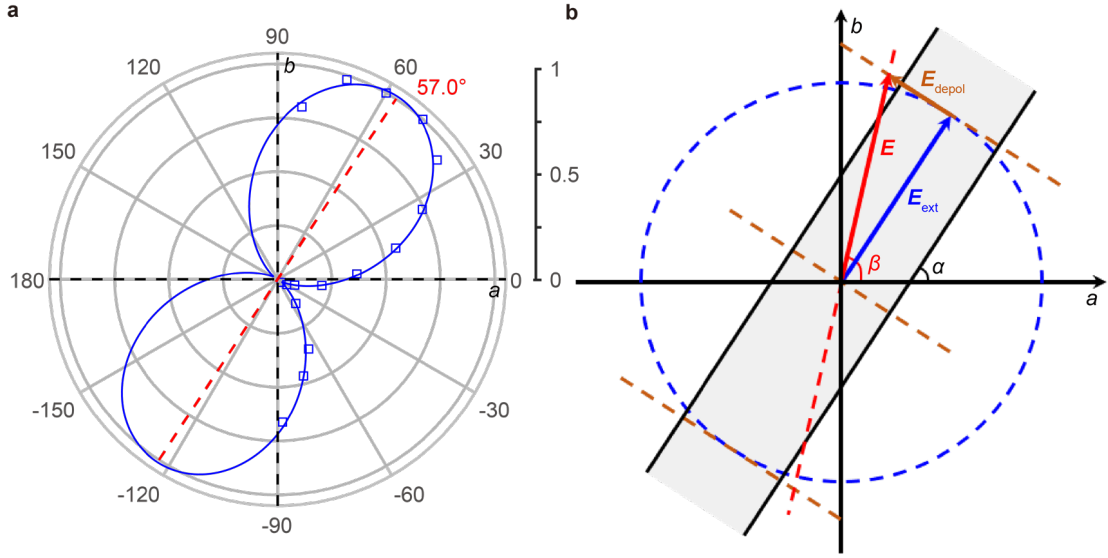

**Fig. S4 | Polarization dependence of the Drude response in skew ribbon arrays. a**, Polarization dependence of the normalized Drude weight of Fig. 2b in the main text. Black dashed lines denote the crystal axes. The parallel direction to ribbons is labeled by the red dashed line with tilted angle of 57°, which is also the light polarization for the maximal Drude weight. **b**, A sketch of electrical fields in a ribbon. The depolarization field $\bm{E}_{\mathrm{depol}}$ is perpendicular to the ribbon edge. The tilted angles of the ribbon and the total electric field $\bm{E}$ are $\alpha = 57°$ and $\beta$, respectively.

### 5. Fitting Details and Errors of LSPRs in Skew Ribbon Arrays.

The fitting details of the spectra in Fig. 2a in the main text are plotted in Fig. S5 and listed in Table S1. The Drude weight $D$, the plasmon intensity $S_{\mathrm{p}}$ and the interband transition intensity $S_{\mathrm{i}}$ are fitting parameters. The plasmon frequency, the interband transition frequency, the Drude and plasmon scattering rates are fixed at 308, 800, 60 and 90 cm$^{-1}$ respectively.

In our experiments, the main sources of error are associated with two parameters: the plasmon resonance frequency $\omega$ and the optimal polarization angle $\phi_{\mathrm{max}}(\omega)$. To determine the error for the first parameter, the raw polarized spectra were fitted using different initial fitting parameters to calculate the average of the resonance frequency. As for $\phi_{\mathrm{max}}$, the error can be attributed to two factors. Firstly, errors arise from the fabrication, such as cutting the skew angle of -33°, and the spectrum acquiring process, such as measuring the polarization angle $\phi$ (with errors of approximately 1°). Secondly, the fitting procedure, which involves fitting the polarization dependence of the plasmon spectral weight by $\cos^2\phi$, introduces a principal error in $\phi_{max}$ of about 1.5°. To obtain an overall estimation of the error in $\phi_{max}$, the contributions from these two factors were combined using the error transfer formula, as shown in Fig. 3a-c in the main text.

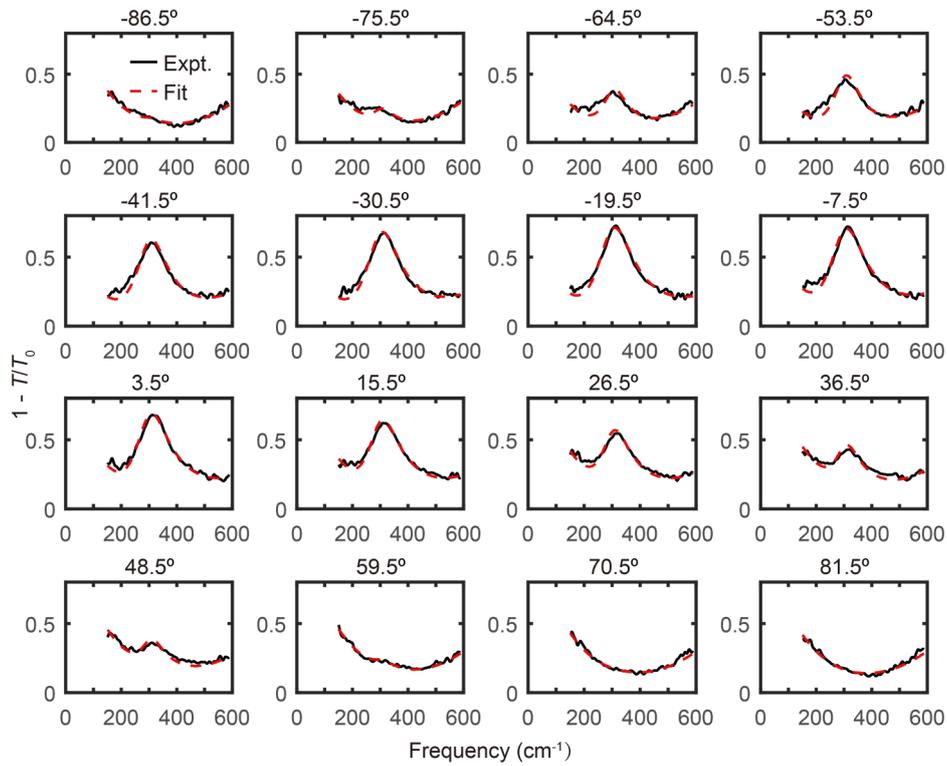

**Fig. S5 | The experimental (solid lines) and fitted (dashed lines) extinction spectra at different polarizations (from -86.5° to 81.5°) from Fig. 2a in the main text.**

| Polar. angle (°) | $D$ ($10^{18}$ a.u.) | $S_p$ ($10^{17}$ a.u.) | $S_i$ ($10^{18}$ a.u.) |
| --- | --- | --- | --- |
| -86.5 | 1.49 | 0.32 | 2.87 |
| -75.5 | 1.40 | 1.20 | 3.09 |
| -64.5 | 1.08 | 3.49 | 2.83 |
| -53.5 | 0.86 | 5.71 | 2.50 |
| -41.5 | 0.79 | 9.55 | 2.07 |
| -30.5 | 0.72 | 11.8 | 1.80 |
| -19.5 | 0.86 | 13.2 | 1.41 |
| -7.5 | 1.03 | 12.8 | 1.69 |
| 3.5 | 1.24 | 11.7 | 1.60 |
| 15.5 | 1.46 | 9.72 | 1.83 |
| 26.5 | 1.71 | 6.97 | 2.24 |
| 36.5 | 1.86 | 4.19 | 2.64 |
| 48.5 | 1.87 | 2.76 | 2.55 |
| 59.5 | 1.87 | 0.58 | 2.96 |
| 70.5 | 1.72 | 0.02 | 2.99 |
| 81.5 | 1.67 | 0.07 | 3.04 |

**Table S1 | Fitting parameters of the spectra in Fig. 2a in the main text.**

## 6. Simulation of Polarized Extinction Spectra in Skew Ribbon Arrays

We simulated the spectra in WTe$_2$ skew ribbon arrays of the skew angle $\theta = -33°$ to verify the equation (1) or (S2). The extinction spectra (denoted as 1-$T$) of five different suspended skew ribbon arrays, with ribbon widths of 0.6, 0.8, 2, 3, 4 μm (different wavevectors) respectively, were calculated using the finite-element software (Comsol Multiphysics) with a refined tetrahedral mesh and the frequency-domain solver of Maxwell equations. The complex dispersive conductivity of WTe$_2$ films (treated as a two-dimensional layer) was determined by previous results[2] (also the following Note 7).

Fig. S6 displays simulations of three representative samples, with resonance frequencies in the elliptic regime (394 cm$^{-1}$), the vicinity of the topological transition point (439 cm$^{-1}$) and the hyperbolic regime (499 cm$^{-1}$), respectively. The corresponding extinction spectra with different incident light polarization (from -90° to 90°, with intervals of 11.25°) are plotted as pseudo color maps in Fig. S6a-c. The angle for maximal plasmon intensity gradually evolves from negative to positive with incremental resonance frequency, demonstrating the clear evidence for the topological transition. Moreover, the Drude response is most intense when the polarization is along the ribbon (57°). These spectra were further fitted by the Drude-Lorentz model to extract the plasmon intensity. The absorption spectra due to the plasmon polariton are plotted in Fig. S6d-f. The extracted plasmon intensity was fitted as a function of $\cos^2\phi$ ($\phi$ is the polarization angle) as shown in Fig. S6g-i. Consistent with equation (1) or (S2), the polarization angle for the maximal plasmon intensity $\phi_{\text{max}}$ is negative (positive) for the elliptic (hyperbolic) topology of the plasmon dispersion and is approximately zero near the transition point. The fitted $\phi_{\text{max}}$ from simulations and the calculated $\phi_{\text{max}}$ by equation (S2) with the corresponding conductivity are listed in Table S2 for quantitative comparison, which agree well with each other (errors originate from the fitting of the simulated spectra). In summary, the simulations verify the theory (Note 2) convincingly.

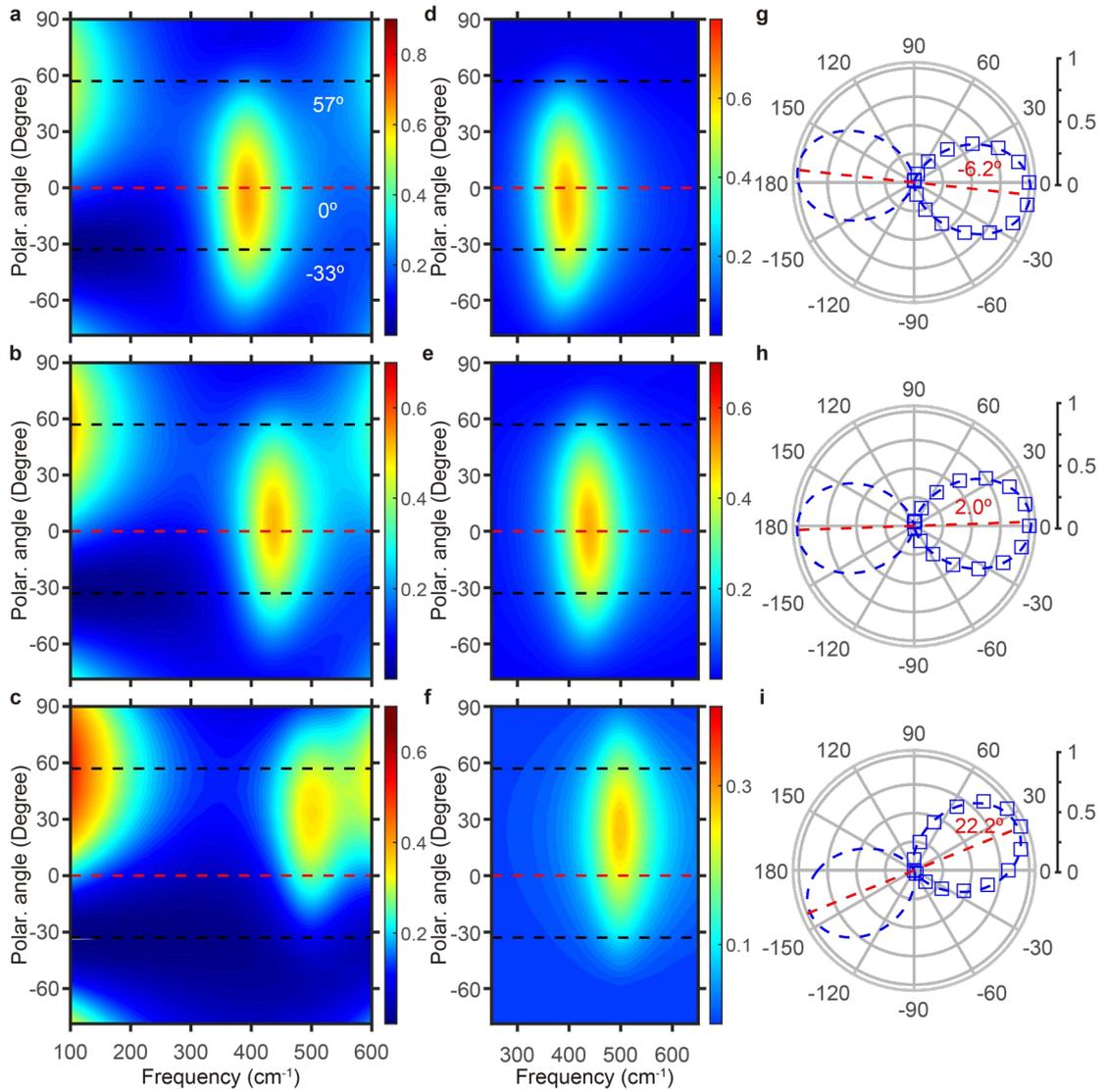

**Fig. S6 | The simulated polarization dependence of plasmon spectra in skew ribbon arrays. a-c,** Polarization dependence of simulated extinction spectra at plasmon frequencies of 394, 439, 499 cm$^{-1}$, respectively. **d-f,** The corresponding pseudo color maps of the absorption spectra due to the plasmon polariton in **a-c**. **g-i,** Polarization dependence of the corresponding normalized plasmon weights of **a-c** or **d-f**. In **a-f**, black dashed lines denote the parallel (57°) and perpendicular (-33°) directions with respect to the ribbon respectively, and the red dashed lines denote *a*-axis.

| Plasmon frequency (cm$^{-1}$) | $\phi_{max}$ (°) by simulations | $\phi_{max}$ (°) by equation (S2) |
|---|---|---|
| 354 | -10.5 ± 0.6 | -10.1 |
| 394 | -6.2 ± 0.5 | -5.7 |
| 439 | 2.0 ± 0.5 | 2.1 |
| 499 | 22.2 ± 0.6 | 22.5 |
| 507 | 26.7 ± 0.7 | 26.6 |

**Table. S2 | The $\phi_{max}$ determined by simulations and calculations of equation (S2).**

### 7. Conductivities Extracted from the Plasmon Dispersion of WTe$_2$.

The detailed fitting process to obtain the optical conductivity of the WTe$_2$ film from the plasmon dispersion has been discussed in the main text of Ref. 2. Here we briefly reiterate. The plasmon resonance frequencies along principal crystal axes were measured in WTe$_2$ rectangular arrays, and the corresponding wavevectors were determined by the structure size. The loss function -Im(1/$\varepsilon$), defined by the imaginary part of the inverse of the dielectric function, was calculated to fit the plasmon dispersion. For the two-dimensional case, the non-local dielectric function is defined as follows,

$$\varepsilon = \varepsilon_{env} + \frac{i\sigma(\omega)}{\varepsilon_0 \omega} \cdot \frac{q}{2} \tag{S6}$$

where $\varepsilon_{env}$ is the substrate permittivity and $q$ is the wavevector (since we consider two principal axes, the vector form of $q$ and tensor form of $\sigma$ are omitted). The conductivities along two principal axes, which are assumed to be a Drude term plus a Lorentz term with parameters to be determined, were obtained by fitting the anisotropic plasmon dispersion using the loss function -Im(1/$\varepsilon$) and equation (S6). The obtained optical conductivities were then substituted into equation (1) (or (S2)) to draw the black solid line in Fig. 3a in the main text.

## 8. Ribbon Width Dependence of LSPRs in Mo$_x$W$_{1-x}$Te$_2$ Skew Ribbons.

In order to observe LSPRs across different frequencies and track the OTT, we changed the ribbon width. Fig. S7 displays the plasmon dispersion of Mo$_x$W$_{1-x}$Te$_2$ ($x$ = 0.278) acquired from skew ribbon arrays ($\theta = -33°$) of different width. The thickness $t$ of the exfoliated film of Mo$_x$W$_{1-x}$Te$_2$ in our paper ranges from 40 to 120 nm. In this thickness range, the electronic band structure remains the same as that of the bulk. The wavevector $q$ in a ribbon array[1] is determined by the ribbon width $L$ (usually from 0.3 to 10 μm) with $q = \pi/L$, which is further normalized to 100 nm thick sample by multiplying $t/100$, since the two-dimensional conductivity is proportional to the film thickness $t$. As shown in Fig. S7, only at low frequencies, the plasmon dispersion follows the $\sqrt{q}$ relation, a characteristic feature for the two-dimensional plasmon polaritons from free carriers[3]. At higher frequencies, the dispersion softens and departs from the $\sqrt{q}$ relation, which is primarily due to the coupling to interband transitions (fitted by $(q/(\varepsilon_{env} + \rho_0 q))^{1/2}$, with $\rho_0$ as a parameter for the screening length )[4,5].

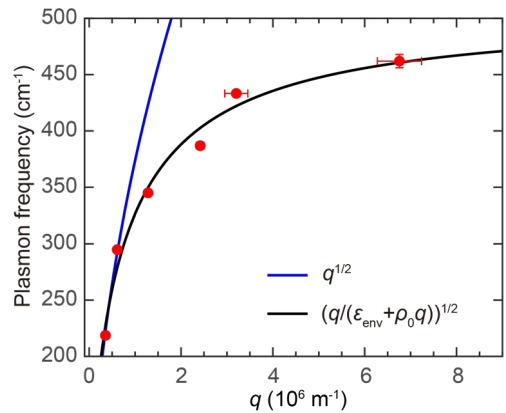

**Fig. S7 | Plasmon dispersion in skew ribbon arrays of Mo$_x$W$_{1-x}$Te$_2$ ($x$ = 0.278).**

## 9. LSPRs in Relatively Large Mo$_x$W$_{1-x}$Te$_2$ Disks.

To investigate the intrinsic intraband plasmon in Mo$_x$W$_{1-x}$Te$_2$, disk arrays with larger diameters (lower wavevectors) were patterned, in which the plasmon frequencies are low enough to stay away from the interband transitions, and the dispersion follows the $\sqrt{q}$ relation. The thickness, disk diameter, plasmon frequencies along *a*- and *b*-axis at different doping are summarized in Table S3 for the samples. The extinction spectra and fitting curves are plotted in Fig. S8. LSPRs along *a*-axis exhibit similar FWHMs for different doping levels, which are 86.6, 93, 102.7 cm$^{-1}$ respectively. The quality factors of those LSPRs are 3.4, 3.1, 2.1 respectively. This suggests that the Drude scattering rates are similar for different doping levels and the doping procedure doesn't degrade the sample quality. Meanwhile, the peak positions of plasmon spectra in Fig. S8 can directly provide us information on the Drude weights, which will be used in Note 10.

| Doping *x* of the sample | 0 | 0.278 | 0.5 |
|---|---|---|---|
| Thickness (nm) | 100 | 50 | 45 |
| Disk diameter (μm) | 4 | 2.92 | 4.6 |
| Plasmon frequency along *a*-axis (cm$^{-1}$) | 295.0 | 291.7 | 214.3 |
| Plasmon frequency along *b*-axis (cm$^{-1}$) | 200.7 | 179.0 | 138.3 |

**Table S3 | Sample parameters of disk arrays at different doping.**

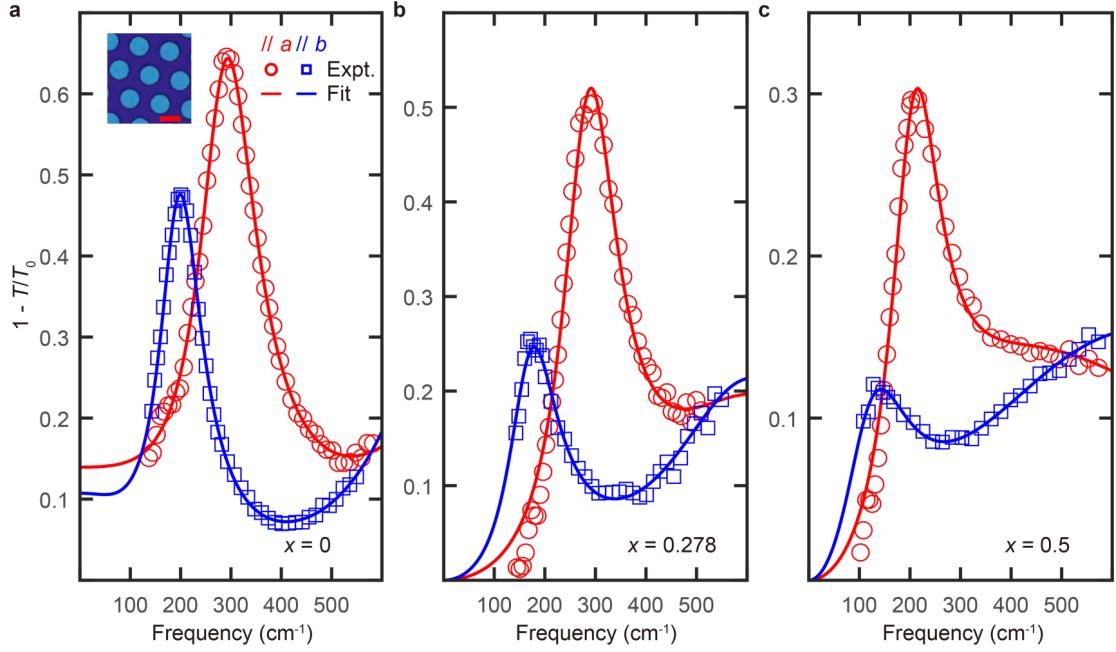

**Fig. S8 | Anisotropic LSPRs of Mo$_x$W$_{1-x}$Te$_2$ ($x$ = 0, 0.278, 0.5) disk arrays.** The upturn of the spectra at low frequency in **a** is due to the Drude absorption from the unpatterned film in the vicinity of the disk array. Inset of **a** shows the optical microscopic image of a disk array at $x$ = 0. Scale bar in red is 5 μm.

### 10. Fitting of the Extinction Spectra of Bare Films and the Extracted Drude Weights.

We fit the extinction spectra of bare films using equations (2) and (3) in the main text. The fitting components in the mid- and far-IR regimes are displayed in Fig. S9, where the first low energy interband transition resonances mentioned in the main text in Fig. 4a-c are labelled by 'L1' in blue. Here, one Drude component (labeled by 'D') and five Lorentz components (labeled by 'L') for $a$-axis (seven Lorentz components for $b$-axis) are included. During the fitting, the Drude scattering rate was fixed at the value derived from the corresponding peak width of the low frequency LSPR.

The Drude response in the lower THz regime is beyond our measurement range, leading to possible underestimation of the Drude weights from the fitting at larger doping (particularly, the 50% doping case). For improvements, at doping composition of $x = 0.5$, we resorted to the plasmon spectra in Note 9, since the Drude feature dictates the plasmon feature, which shifts into our measurement range. The plasmon frequency in the small $q$ limit ($\omega \propto \left(q \cdot \frac{n_e}{m_e}\right)^{1/2}$, $n_e$ is the sheet carrier density) at different doping is related to the Drude weight[3] ($D = \frac{\pi e^2 n_e}{m_e}$). Therefore, at known thickness and plasmon wavevector, $D_{0.5}$ (0.5 is the composition ratio) can be calculated according to the peak frequencies of Fig. S8. Specifically, by comparing to the fitted $D_0$ of WTe$_2$, the Drude weight $D_{0.5}$ can be obtained. By equation (3), the DC conductivity $\sigma_{0.5}(0)$ is equal to $\frac{D_{0.5}}{\pi \Gamma}$, which determines the extinction $A_{0.5}$ at zero frequency according to equation (2). It was included to fit the extinction spectrum of the unpatterned film with $x = 0.5$. Finally, the fitted thickness-normalized Drude weights ($D$ divided by the thickness $t$, namely, the Drude weight for the bulk) for $x = 0.5$ and those obtained through direct fitting of the film spectra for $x = 0$, and $x = 0.278$ are displayed in Fig. S10.

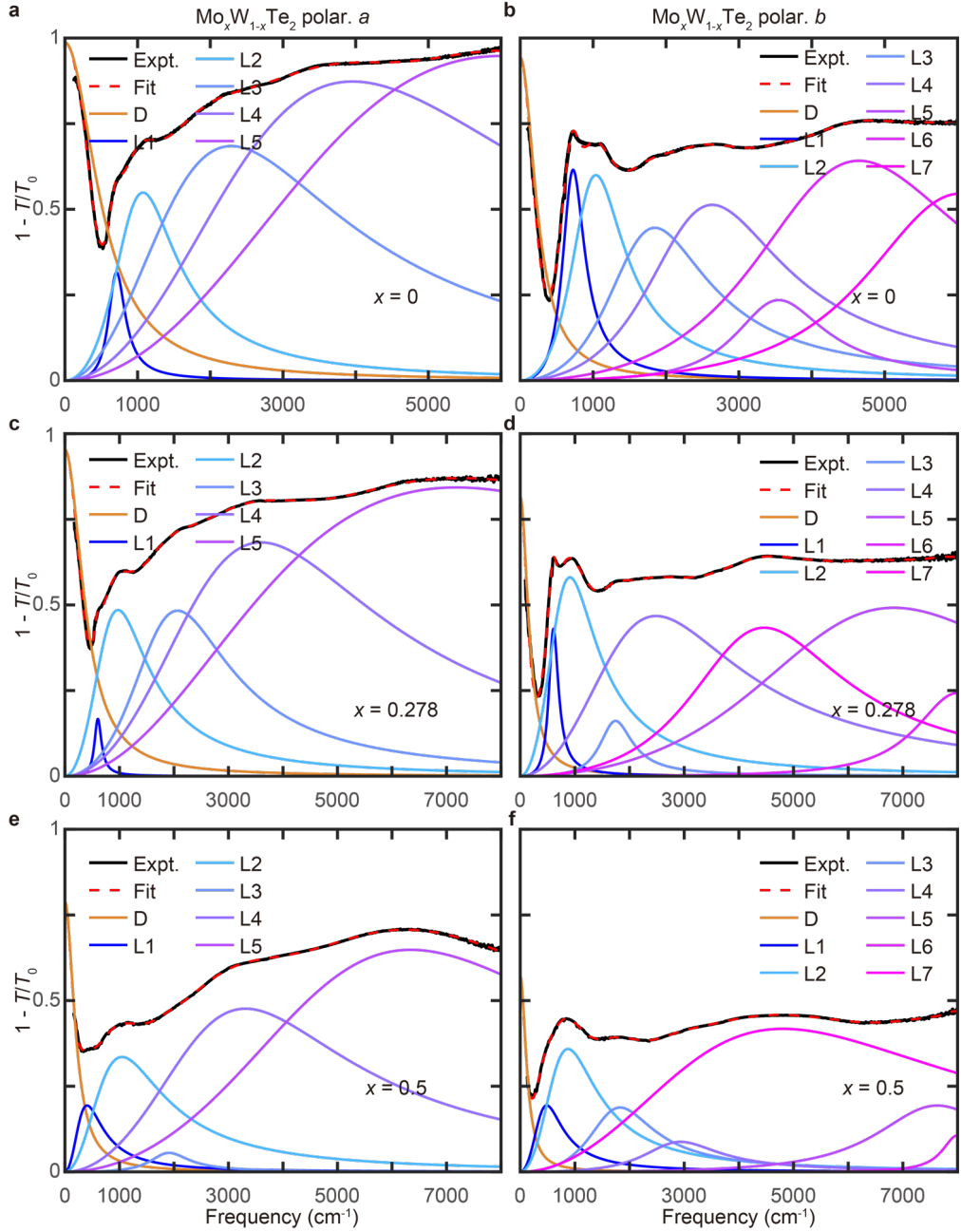

**Fig. S9 | Drude-Lorentz model fittings of extinction spectra in Figs. 4a-c in the main text (here with extended frequency range).** The left and right panels are for *a*-axis and *b*-axis polarizations respectively. The colored solid lines labelled by "D" (Drude response) or "L" (interband transitions) are Drude and Lorentz components, which were obtained by substituting the corresponding conductivity components into equation (2) in the main text.

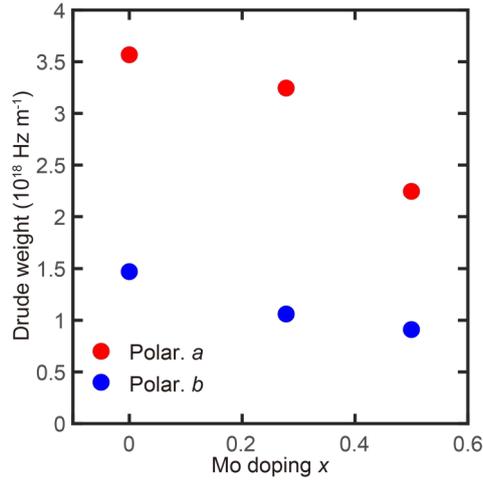

**Fig. S10 | Extracted Drude weights of $Mo_xW_{1-x}Te_2$ ($x$ = 0, 0.278, 0.5) films from Fig. S9, normalized by the thickness.**

## 11: IFCs of Plasmon Dispersion in $Mo_xW_{1-x}Te_2$.

The loss function $\text{Im}(-1/\varepsilon)$ (the imaginary part of the inverse of the dielectric function) is calculated in the two-dimensional wavevector space to plot the IFCs of the plasmon dispersion in $WTe_2$ at different Mo doping levels and temperatures. This is achieved by utilizing equation (S6) and the fitted optical sheet conductivity obtained from experimental data (e.g., Fig. 4d-f and Fig. S13b).

Fig. S11 presents pseudocolor maps of the calculated loss function, showcasing the evolution of the IFCs in $Mo_xW_{1-x}Te_2$ (a for $WTe_2$, b for 27.8% Mo doping, c for 50% Mo doping). The solid black curves in Fig. S11 depict the fitted dispersion (IFCs). In particular, Fig. S11a illustrates the gradual transition of the IFCs from an elliptic shape to a hyperbolic shape as the plasmon resonance frequency increases. The evolution signifies the OTT of the IFCs of plasmon polaritons in $WTe_2$. Additionally, the redshift of the OTT energy with increasing Mo doping is evident, which aligns with the findings from our polarization experiments.

Using the same method, the temperature dependence of the IFCs at a specific frequency (440 cm$^{-1}$) in WTe$_2$ is also examined, as depicted in Fig. S12. As the temperature increases, the IFCs undergo a gradual transition from a hyperbolic shape to an elliptic shape. This observation is in line with the blueshift of the energy of the OTT at higher temperatures, as illustrated in Fig. S13b.

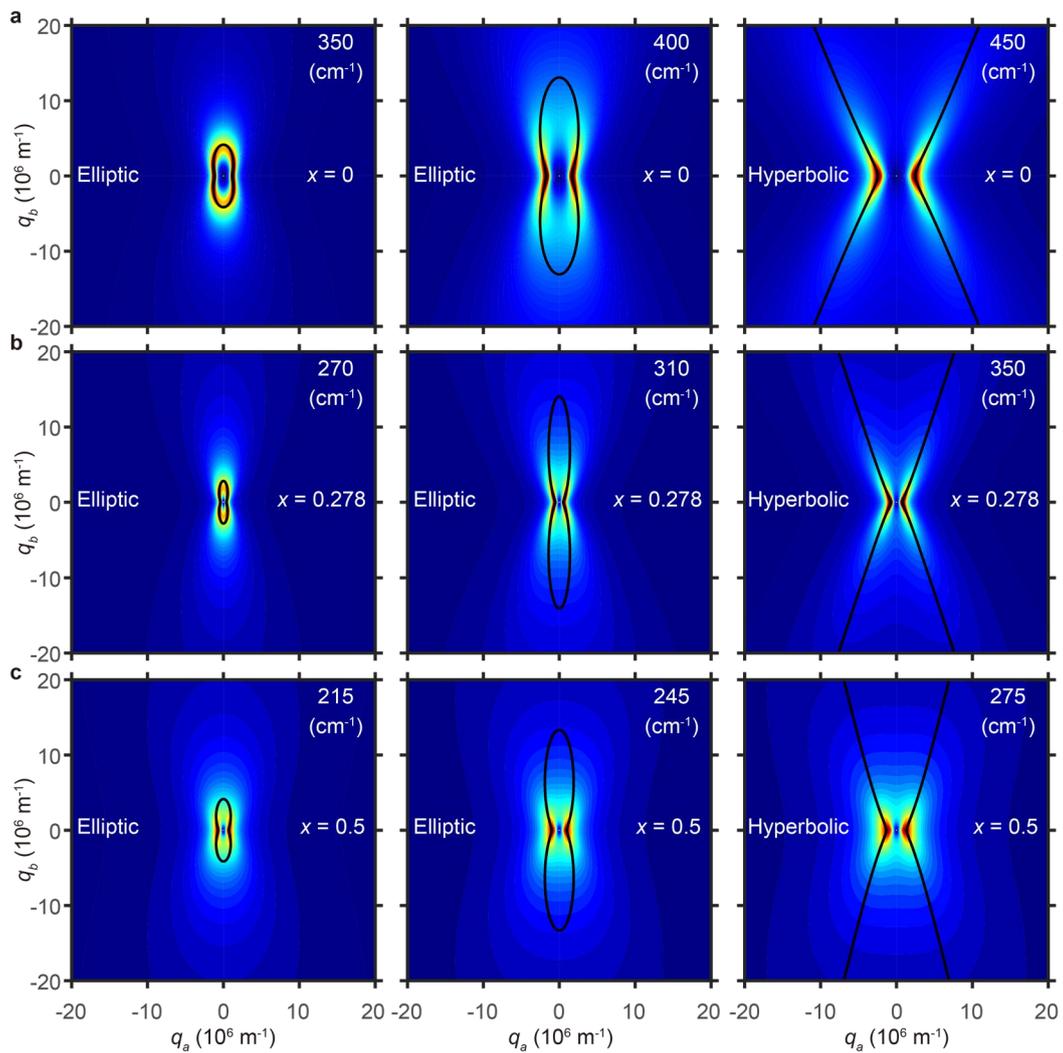

**Fig. S11 | Mo doping effect on the IFCs of plasmon dispersion at different resonance frequencies in WTe$_2$.** The Mo doping $x$ is 0, 0.278, 0.5 in **a**, **b** and **c** respectively. The loss function is displayed as pseudocolor maps in the two-dimensional wavevector space. The solid black curves represent the fitted dispersion (IFCs) of plasmon polaritons.

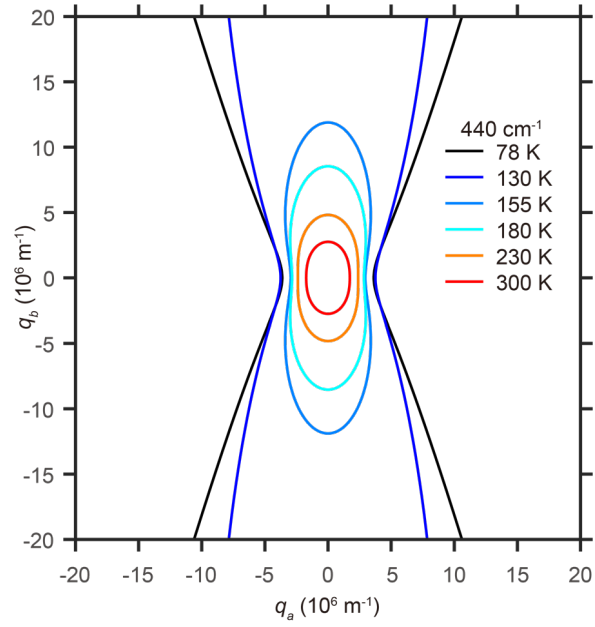

**Fig. S12 | Temperature dependence of the IFCs of plasmon dispersion in WTe$_2$.** The solid curves are fitted to the calculated loss functions at each temperature. The plasmon resonance frequency is 440 cm$^{-1}$.

**12. Temperature Dependence of the OTT in WTe$_2$.**

The extinction spectra and their fitting curves (here with extended frequency range) of a WTe$_2$ bare film (35 nm thickness), the extracted imaginary parts of the optical conductivities along both axes, and the polarization angle $\phi_{max}$ as a function of the plasmon frequency at different temperatures are displayed in Fig. S13.

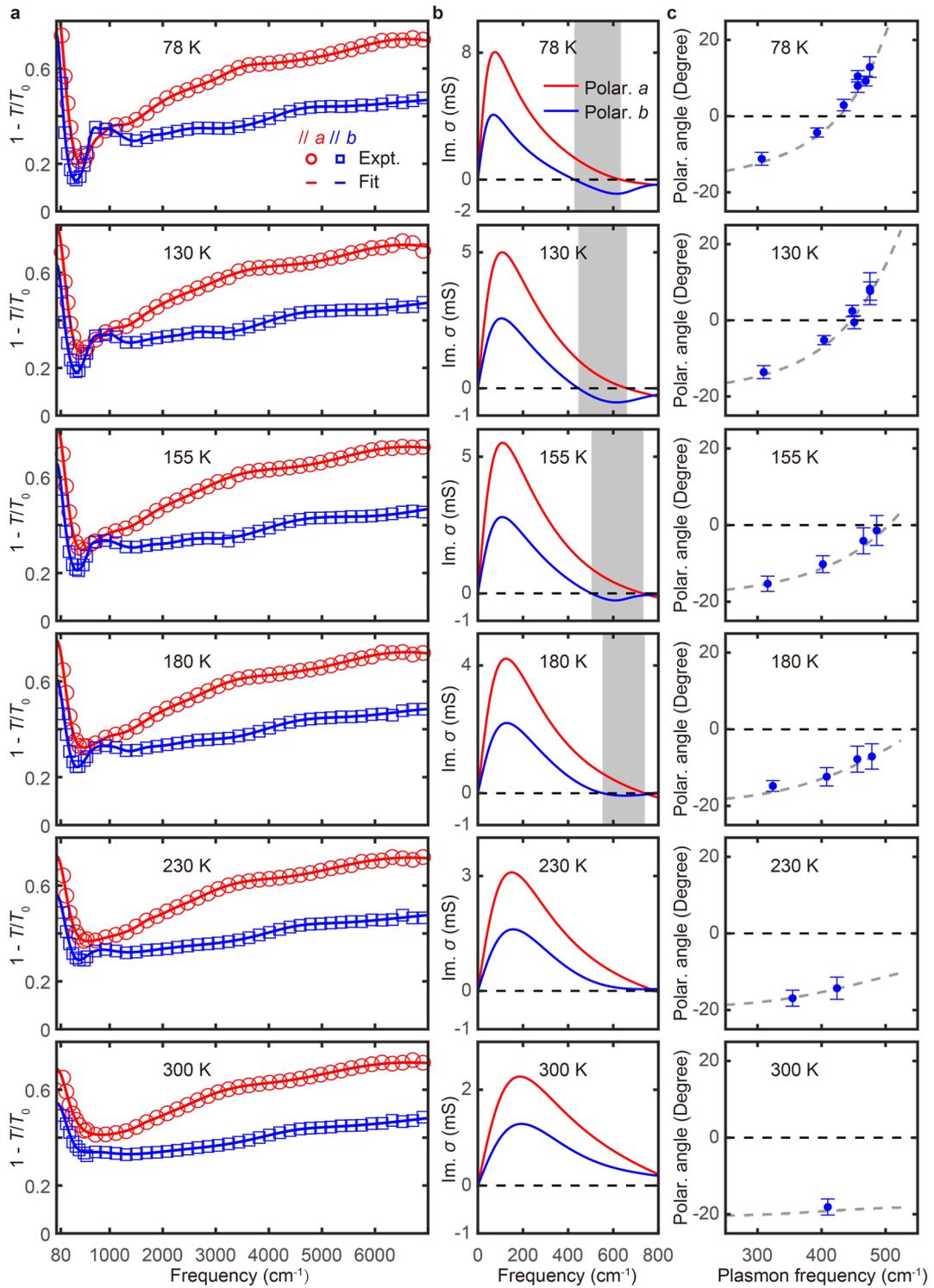

**Fig. S13 | Temperature dependence of OTTs in WTe$_2$. a,** The anisotropic extinction spectra and fitting curves of a bare film of WTe$_2$ (here with extended frequency range). The sample was maintained at temperatures from 78 K to 300 K from the top to the bottom panel. **b,** The corresponding extracted imaginary parts of the optical conductivities along two crystal axes in **a,** with the hyperbolic regime shaded. **c,** The optimal polarization $\phi_{max}$ as a function of the plasmon frequency at the corresponding temperature.